\documentclass[sigconf]{acmart}

\usepackage{threeparttable}
\usepackage{multirow}
\usepackage{subfigure}
\usepackage{caption}
\usepackage{makecell}
\usepackage{balance}
\usepackage{graphicx}
\usepackage{epstopdf}
\usepackage{algorithm}
\usepackage{algorithmic}
\usepackage{adjustbox}
\usepackage{xcolor}
\usepackage{listings}
\usepackage{pifont}
\usepackage{xcolor}
\lstdefinestyle{mystyle}{
	moredelim=**[is][\color{red!100}]{@}{@},
}
\renewcommand\footnotetextcopyrightpermission[1]{}

\begin{document}

\title{HyperGo: Probability-based Directed Hybrid Fuzzing}

\author{Peihong Lin}
\affiliation{%
 \institution{National University of Defense Technology}
 \streetaddress{Kaifu Qu}
 \city{Changsha}
 \country{China}}
\email{phlin22@nudt.edu.cn}

\author{Pengfei Wang}
\affiliation{%
 \institution{National University of Defense Technology}
 \streetaddress{Kaifu Qu}
 \city{Changsha}
 \country{China}}
\email{pfwang@nudt.edu.cn}

\author{Xu Zhou}
\affiliation{%
 \institution{National University of Defense Technology}
 \streetaddress{Kaifu Qu}
 \city{Changsha}
 \country{China}}
\email{zhouxu@nudt.edu.cn}

\author{Wei Xie}
\affiliation{%
 \institution{National University of Defense Technology}
 \streetaddress{Kaifu Qu}
 \city{Changsha}
 \country{China}}
\email{xiewei@nudt.edu.cn}

\author{Kai Lu}
\affiliation{%
 \institution{National University of Defense Technology}
 \streetaddress{Kaifu Qu}
 \city{Changsha}
 \country{China}}
\email{kailu@nudt.edu.cn}

\author{Gen Zhang}
\affiliation{%
 \institution{National University of Defense Technology}
 \streetaddress{Kaifu Qu}
 \city{Changsha}
 \country{China}}
\email{zhanggen@nudt.edu.cn}

%

\begin{abstract}

Directed grey-box fuzzing (DGF) is a target-guided fuzzing intended for testing specific targets (e.g., the potential buggy code). Despite numerous techniques proposed to enhance directedness, the existing DGF techniques still face challenges, such as taking into account the difficulty of reaching different basic blocks
when designing the fitness metric,
and promoting the effectiveness of symbolic execution (SE) when solving the complex constraints in the path to the target. 
In this paper, we propose a directed hybrid fuzzer called HyperGo. To address the challenges, we introduce the concept of path probability and combine the probability with distance to form an adaptive fitness metric called \textit{probability-based distance}.
By combining the two factors, probability-based distance can adaptively guide DGF toward paths that are closer to the target and have more easy-to-satisfy path constraints. 
Then, we put forward an Optimized Symbolic Execution Complementary (OSEC) scheme to combine DGF and SE in a complementary manner. The OSEC would prune the unreachable branches and unsolvable branches, and prioritize symbolic execution of the seeds whose paths are closer to the target and have more branches that are difficult to be covered by DGF. 
We evaluated HyperGo on 2 benchmarks consisting of 25 programs with a total of 120 target sites. The experimental results show that HyperGo achieves 37.75$\times$, 29.11$\times$, 23.34$\times$, 95.61$\times$ and 143.22$\times$ speedup compared to AFLGo, AFLGoSy, BEACON, WindRanger, and ParmeSan, respectively in reaching target sites, and 3.44$\times$, 3.63$\times$, 4.10$\times$, 3.26$\times$, and 3.00$\times$ speedup in exposing known vulnerabilities. Moreover, HyperGo discovered 10 undisclosed vulnerabilities from 5 real-world programs. 
\end{abstract}





\maketitle

\section{Introduction}
Grey-box fuzzing has been a scalable and effective approach to discovering vulnerabilities in software in recent years \cite{1,2,3,4}. Based on the feedback information from the execution of the program under test (PUT), grey-box fuzzers utilize an evolutionary algorithm to generate specific inputs that can cause erroneous runtime behavior (e.g., memory corruptions or data abort) of PUT. Most existing fuzzers are coverage-guided (CGF) \cite{5,6,7,8} as they focus on improving the code coverage to test the deeper level of code. However, not all parts of the code in PUT are equally important because the majority of the code are safe and only a small portion has vulnerabilities. For example, according to Shin \textit{et al}. \cite{9}, only 3$\%$ of the source code files in Mozilla Firefox have vulnerabilities. Thus, researchers aim to focus on strengthening the tests for the vulnerable parts of the code.
To achieve directedness, the originally directed fuzzers were based on symbolic execution (SE) \cite{10,11,12,13}, which uses program analysis and constraint solving to generate inputs that exercise different program paths. Such directed fuzzers cast the reachability problem as an iterative constraint satisfaction problem. However, since directed symbolic execution relies on heavyweight program analysis and constraint solving, it suffers from scalability and compatibility limitations \cite{51}. 

In 2017, a directed grey-box fuzzer AFLGo \cite{14} was proposed. It leverages lightweight compile-time instrumentation to drive the fuzzing toward a set of pre-defined target locations. Different from CGF which strives to increase the code coverage, DGF intends to reach and test the target sites (e.g., the potential buggy code). 
Based on the call graph and control-flow graph information of the PUT, DGF uses the distance between inputs and target sites as the fitness metric to assist seed selection and seed energy allocation. Thus, DGF can prioritize the seeds that are more likely to reach the targets (i.e., optimal seeds), 
which makes DGF outperforms CGF in specific scenarios, such as patch testing \cite{1dVul}, bug reproduction \cite{16,17,18}, and potential buggy code verification \cite{19,20}.
To accelerate reaching targets and exposing vulnerabilities, the state-of-the-art DGF techniques proposed in recent years have employed various methods to enhance directedness. For instance, some DGF techniques redefine the fitness metric based on trace similarity (Hawkeye \cite{21}), data-flow graph (CAFL \cite{22}, WindRanger \cite{23}), and sequence coverage (LOLLY \cite{24}) while some other DGF techniques prune infeasible paths (BEACON \cite{26}), use the number of oracle queries required by a fuzzing algorithm to find a target-reaching input ($MC^2$ \cite{MC}) and utilize symbolic execution to penetrate the complex path constraints toward targets, namely directed hybrid fuzzing \cite{28,29,30}. However, despite these achievements, the existing DGF techniques still face two challenges.

\textbf{Challenge 1: Taking into account the difficulty of reaching different basic blocks to design a more effective fitness metric.}
Following AFLGo, most of the state-of-the-art DGF techniques proposed new fitness metrics and methods based on the knowledge of program analysis (e.g., DBB-distance in WindRanger and constraint-distance in CAFL). Although the methods can be beneficial in testing certain programs, they may be inaccurate while some specific programs do not meet their assumptions (e.g., WindRanger assumes that the complexity of path constraints is related to the number of corresponding seed bytes). During the fuzzing process, reaching different basic blocks has different probabilities since the path constraints are not all equally satisfied. It is challenging for the fuzzer to reach target sites through those basic blocks that are difficult to cover. Thus, it is a challenge to adaptively analyze the probability of reaching different basic blocks without prior knowledge of program analysis and combine the probability with the basic-block-level distance (i.e., BB distance) to form a more effective fitness metric.

\textbf{Challenge 2: Promoting the effectiveness of symbolic execution when solving the complex constraints in the path to the targets.}
As DGF takes the random mutation, it may not be able to satisfy the complex constraints within the allotted time budget. To address this issue, a complementary symbolic execution technique can be introduced to assist DGF, which should meet three requirements: (1) preferentially solving the path constraints of the branches closer to target sites, (2) preferentially solving the path constraints of the branches that are difficult to be covered by DGF, and (3) pruning the branches that are unsolvable by symbolic execution or do not contribute to reaching the target. 
However, the symbolic execution techniques used in state-of-the-art hybrid fuzzers (such as SAVIOR \cite{28}, Symcc \cite{40}, DigFuzz \cite{34}, and Hydiff \cite{29}) cannot meet all three requirements simultaneously. Thus, it is a challenge to design a complementary symbolic execution technique to effectively assist DGF.

In this paper, we propose HyperGo, the probability-based directed hybrid fuzzing.
For challenge 1, we introduce the concept of path probability which is dynamically calculated based on branch hits, and then combine the path probability with BB distance to form an adaptive fitness metric called \textbf{\textit{probability-based distance}}. The path probability reflects the difficulty of DGF reaching one basic block while the BB distance reflects the likelihood of DGF reaching the target sites through the basic block.
By combining the two factors, probability-based distance can effectively guide DGF toward paths that are closer to the target and have easier-to-satisfy path constraints (Section \ref{section:distance}). Upon the introduction of a new fitness metric, it is imperative to optimize the power schedule to adaptively balance the exploitation of seeds with short distances and the exploration of more seeds that are reachable to the target sites (i.e., reachable seeds).
To achieve this, we develop an optimization strategy for the power schedule called the Directed Multi-Armed Bandit (DMAB) model. Based on the continuously changing probability-based distance and path probability,  the DMAB model adaptively assigns more energy to seeds that have shorter seed distances and higher probabilities of covering new branches.

For challenge 2, we propose an Optimal Symbolic Execution Complementary (i.e., OSEC) scheme that combines DGF and SE in a complementary way. In OSEC, we implement three strategies to improve the effectiveness of the combination between DGF and SE: (1) pruning branches that do not contribute to reaching target sites (i.e., unreachable branches), (2) pruning branches whose path constraints cannot be solved by SE (i.e., unsolvable branches), and (3) prioritizing the symbolic execution of the seeds whose paths are closer to the target and have more branches that are difficult to be covered  by DGF. The first and second strategies aim at improving the efficiency of SE, while the third strategy aims at creating complementarity between SE and DGF. Specifically, we prompt DGF to explore branches with simpler path constraints, while SE is geared towards solving the path constraints of more complex branches. Based on this method, DGF and SE work in a complementary way and reach target sites more efficiently.

The main contributions of this paper are summarized as follows:

\begin{itemize}

\itemsep -0.3cm 

\item  We propose an adaptive fitness metric called probability-based distance, which combines basic-block-level distance with path probability to achieve higher accuracy and efficiency. It can steer DGF to reach the target sites faster through the closer paths which are easier to re-exercise.

\itemsep -0.1em 

\item We propose a power scheduling optimization strategy called the DMAB model to implicitly balance the exploitation of seeds with short distances and the exploration of more reachable seeds. The seeds that have shorter seed distances and higher probabilities of covering new branches will be assigned more energy.

\itemsep -0.1em 

\item  We propose an OSEC scheme to combine DGF and SE in a complementary manner. The OSEC prunes the unreachable and unsolvable branches and prioritizes the symbolic execution of the seeds whose paths are closer to the target and have more branches that are difficult to be covered  by DGF.

\itemsep-0.1em 

\item  We implemented a tool named HyperGo and evaluate it on 2 datasets consisting of 25 programs with a total of 120 target sites.
The experimental results show that HyperGo achieves 37.75$\times$, 29.11$\times$, 23.34$\times$, 95.61$\times$ and 143.22$\times$ speedup compared to AFLGo, AFLGoSy, BEACON, WindRanger, and ParmeSan, respectively in reaching target sites, and 3.44$\times$, 3.63$\times$, 4.10$\times$, 3.26$\times$, and 3.00$\times$ speedup in exposing known vulnerabilities. Moreover, HyperGo discovered 10 undisclosed vulnerabilities from 5 real-world programs.

\itemsep -0.1em 

\item  HyperGo is publicly available on our website. \url{https://gitee.com/paynelin/hypergo}

\end{itemize}

\section{Background and Motivation}

\subsection{Background}
\label{background}
We first introduce the background knowledge of the DGF techniques and hybrid fuzzing techniques.

\textbf{Distance calculation and power schedule.} AFLGo calculates the distances between the inputs and predefined targets. The seed distance is calculated as the arithmetic mean of BB distances of the basic blocks in the seed's trace. The BB distance is determined by the number of edges in the call graph and control-flow graphs to the target basic blocks while each edge has the same weight. Then, at run-time, AFLGo views the fuzzing process as a Markov chain and leverages a simulated annealing strategy to gradually assign more energy to the seeds that are closer to targets. It casts reachability as an optimization problem to minimize the distance between the generated seeds and their targets.

\textbf{Hybrid fuzzing.} Hybrid fuzzing involves a combination of fuzzing and symbolic execution. Fuzzing is excellent at exploring common code regions and discovering more paths, while symbolic execution can track seed execution paths and reverse branch conditions to identify the branches that are not covered by fuzzing, namely \textit{unexplored branches}. The constraint-solver is then invoked to solve the path constraints of the unexplored branches in the abstracted syntax and generate new seeds to assist fuzzing in satisfying the path constraints. However, existing hybrid fuzzing techniques are not well-suited to meet the needs of DGF, and they face four problems: (1) solving unreachable branches, (2) solving unsolvable branches, (3) solving the branches that had been covered by DGF, and (4) performing not well in adaptively adjusting the solving priority of seeds and the time budget of solving branches. These problems are more pronounced in DGF compared to CGF.

\subsection{Motivation }
\label{motivation}	
\begin{figure}[b]
	\setlength{\abovecaptionskip}{0.cm}
	\centering
	\noindent \includegraphics[width=0.8\columnwidth]{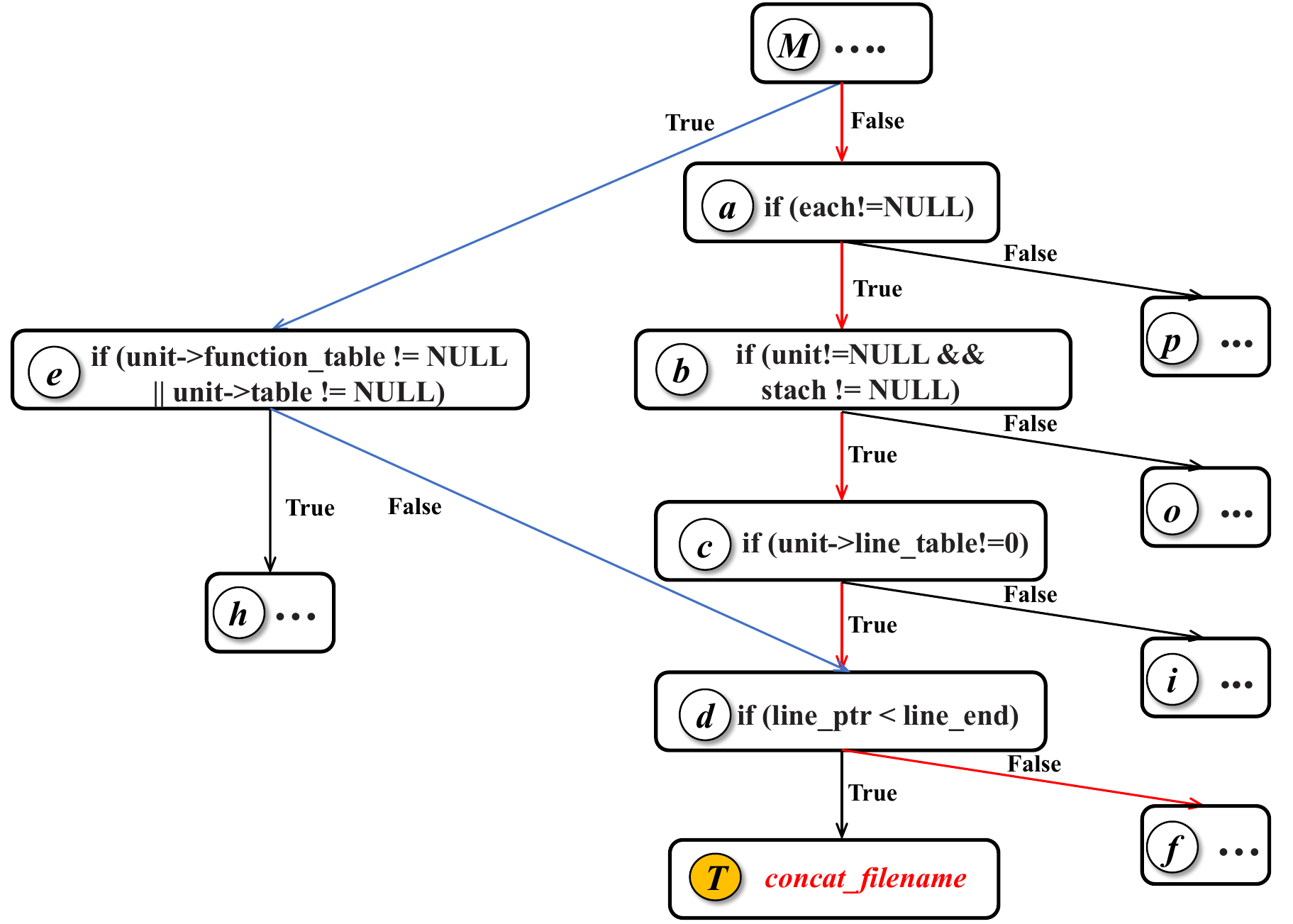}
	\caption{\label{fig:1}  Two execution traces toward target function \texttt{concat\_filename()}. The nodes denote the basic blocks, and the branch conditions are represented nearby.}
\end{figure}

\textbf{Example of challenge 1.} Figure \ref{fig:1} shows a real-world example (CVE-2017-15023) in GNU Binutils 2.29 \cite{GNU}. Two execution traces (Trace 1 $<$\textit{M}, \textit{a}, \textit{b}, \textit{c}, \textit{d}, \textit{f}$>$ is marked as red lines, and Trace 2 $<$\textit{M}, \textit{e}, \textit{d}, \textit{f}$>$ is marked as blue lines) are toward the bug function \texttt{concat\_filename()}.
The call site of \texttt{concat\_filename()} is denoted as \textit{T}. 
Following AFLGo, the seed distance of the seed covering Trace 1 (i.e., Seed 1) should be (1+2+3+4+3)/5=2.3, and the seed distance of the seed covering Trace 2 (i.e., Seed 2) should be (1+2+3)/3=2. Thus, Seed 2 has a shorter seed distance and will be assigned more energy. WindRanger introduces the concept of \textit{NumOfEffectiveBytes} and combines it with the seed distance to form a new fitness metric called \textit{DBB-distance}. Based on the new fitness metric, Seed 1 still has a greater DBB-distance (2.18) than Seed 2's DBB-distance (2.06). Thus, in both works, Seed 2 is given priority.

\graphicspath{{images/} }

However, due to the fuzzer's random mutation strategy, it is difficult for the fuzzer to satisfy this branch condition and simultaneously cover the branches $<$\textit{e}, \textit{d}$>$ and $<$\textit{d}, \textit{T}$>$. Therefore, even though Trace 2 is closer to the target site based on static analysis, it is infeasible for the fuzzer to reach the target sites. Even if Seed 2 is given more energy, the fuzzer still struggles to mutate Seed 2 and reach the target sites.
Notably, both BEACON \cite{26} and SelectFuzz \cite{selectfuzz} would perform poorly in inferring path feasibility in this case. The symbolic execution of BEACON would fail to recognize the path infeasibility of Trace 2 due to the complex path constraints, and SelectFuzz \cite{selectfuzz} would fail to recognize that Trace 2 is more complex than Trace 1 based on the number of successor basic blocks. Based on this real-program-based example,\textbf{ we can conclude that the static fitness metric based on program analysis may be inaccurate. We need an adaptive fitness metric that combines the probability to more accurately guide DGF in different real programs and different fuzzing stages.}

\textbf{Example of challenge 2.} During our research, we investigated the state-of-the-art directed hybrid fuzzers and evaluated their performance in directed testing. 
For instance, we used SAVIOR \cite{28} to test tcpdump five times, and each test lasted for 24 hours. According to the test results, we found that only \textbf{37\%} of the new inputs generated by the symbolic executor are reachable, only \textbf{28\%} of the attempts to generate new seeds are successful, and only \textbf{41\%} of the newly generated seeds are regarded as interesting. Furthermore, SAVIOR is incapable of dynamically adjusting the time budget for solving different branches, resulting in the skipping of some important branches within a very limited time budget and the inability to perform symbolic execution on all seeds within 24 hours. The investigation of SAVIOR demonstrates the issues of the state-of-the-art directed hybrid fuzzing techniques, such as solving unreachable or unsolvable branches, and generating a low proportion of interesting seeds. 
\textbf{Thus, we need to redesign the working scheme of symbolic execution to alleviate these issues and combine DGF and SE in a complementary manner.}

\vspace{-0.2cm}
\section{Probability-based Directed Hybrid Fuzzing}

%
\graphicspath{{images/} }
\begin{figure*}[tp]
\vspace{-0.3cm}
\setlength{\abovecaptionskip}{0em}
\centering
\includegraphics[width=17.8cm,height=3.8cm]{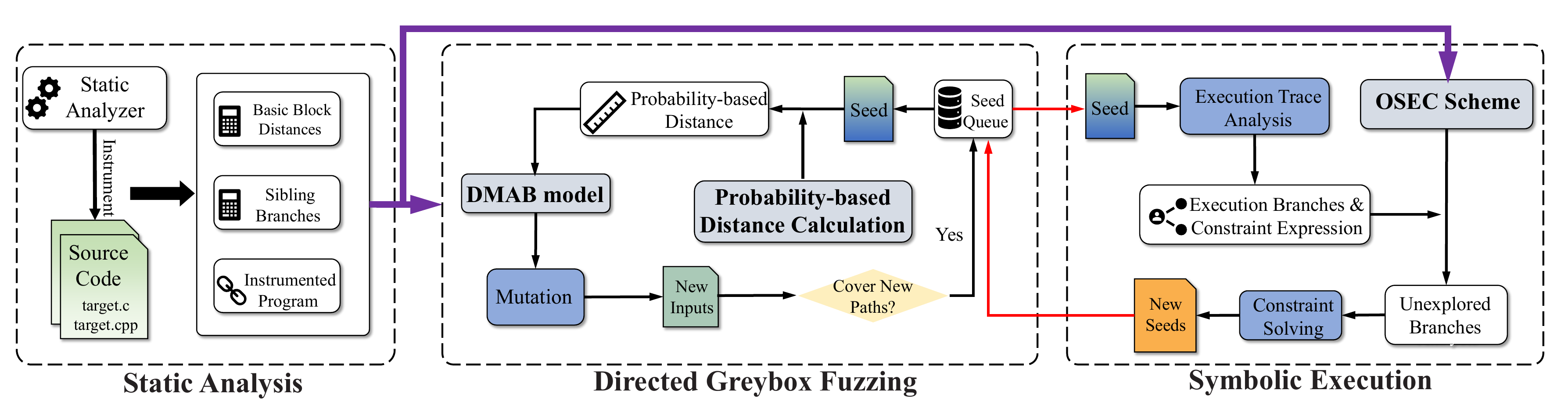}
\caption{\label{fig:2}  The overview of HyperGo. }
\vspace{-1.5em} 
\end{figure*}

In this paper, we propose a probability-based directed hybrid fuzzer named HyperGo. 
As Figure \ref{fig:2} shows, HyperGo consists of the following three major components. 

\textbf{Static analyzer.} The static analyzer is designed to provide precise information to both the directed greybox fuzzer and the symbolic executor, including unique basic block addresses, BB distances, and sibling branches of each branch. To calculate the address of all basic blocks, the static analyzer utilizes a hash algorithm based on the last statements of each basic block (for example, \textit{exam.cpp:24} indicates line 24 of the file exam.cpp). Then, the static analyzer identifies sibling branches based on the successive basic blocks of each basic block. For example, if a basic block $B_1$ has two successors, $B_2$ and $B_3$, the branch $<$$B_1, B_2$$>$ and branch $<$$B_1, B_3$$>$ are sibling branches. Additionally, we adopt the same method as AFLGo to calculate the BB distances.

\textbf{Directed greybox fuzzer.} The directed greybox fuzzer continuously mutates seeds in an attempt to generate inputs that can cover target sites.
We introduce the probability-based distance calculation module and the DMAB model to the fuzzer. The calculation module calculates the probability-based distance by analyzing the statistical path probability and BB distance, and the DMAB model optimizes the power schedule based on this new fitness metric.

\textbf{Symbolic executor.} The symbolic executor tracks the path of the seeds provided by the directed greybox fuzzer to identify unexplored branches. Then, the symbolic executor invokes the constraint-solver to solve the path constraints of the unexplored branches. We introduce the OSEC scheme to alleviate the limitations of hybrid fuzzing and complement the combination of DGF and SE.

At compile time, the static analyzer analyzes the program and stores the analysis information, such as BB distance, locally. This information is loaded by the directed greybox fuzzer and the symbolic executor as the fuzzing campaign is launched. During the fuzzing process, the directed greybox fuzzer continuously generates seeds and provides them to the symbolic executor. The symbolic executor tracks the paths of these seeds to identify unexplored branches and generates new inputs by solving path constraints of the unexplored branches. The new inputs are then fed back to the directed greybox fuzzer, enabling it to quickly reach target sites.

\subsection{Probability-based Distance}
\label{section:distance}
We introduce the concept of path probability and combine it with BB distance to form an adaptive fitness metric called\textbf{\textit{ probability-based distance}}. In this section, we provide a detailed introduction to the concept and calculation method of probability-based distance.

\subsubsection{Definition of Probability-based Distance}
In recent works, such as MDPC \cite{33}, a seed's execution trace in fuzzing is treated as a Markov Chain, and the path probability is calculated based on branch probabilities in the execution path. Based on this assumption, we propose a definition for path and path probability:

\textbf{Definition 1 (The path of a basic block)} Given the execution trace $Trace$ for a seed and a basic block $m$ in $Trace$, the path of $m$, denoted as $path_m$, is a sequence of basic blocks in $Trace$ from the entry basic block to $m$.

\textbf{Definition 2 (Path Probability)} The path probability of a basic block $m$, given its path $path_m$, is defined as the product of the probabilities of the branches that are covered by $path_m$:
\begin{equation}
	\label{eq2}
	P(path_m) = \prod {\{ P(b{r_i})|b{r_i} \in path_m\} } 
\end{equation}

Where $br_i$ denotes a branch in $path_m$, $P(path_m)$ and $P(br_i)$ denote the path probability and branch probability, respectively. The branch probability is calculated based on the branch hits:

\begin{equation}
	\label{eq3}
	P(b{r_j}) = {\frac{{hi{t_j}}}{{\sum\limits_{k = 1}^{total} {(hi{t_k})} }}}
\end{equation}
Where $P(br_j)$ denotes the branch probability of the $j^{th}$ branch of the branch condition ($j$ can be 1 or 2 for a binary branch), $total$ denotes the total number of branches, and $hit_j$ denotes the number of branch hits of the $j^{th}$ branch. 
Notably, the branch hits of the branches that are not covered by fuzzing will be regarded as 1 since we believe that any branch has the probability of being discovered.
\graphicspath{{images/}}

\begin{figure}[hpb]
\vspace{-0.1cm}
	\setlength{\abovecaptionskip}{0.1cm}
	\centering
	\noindent \includegraphics[width=7cm,height=5cm]{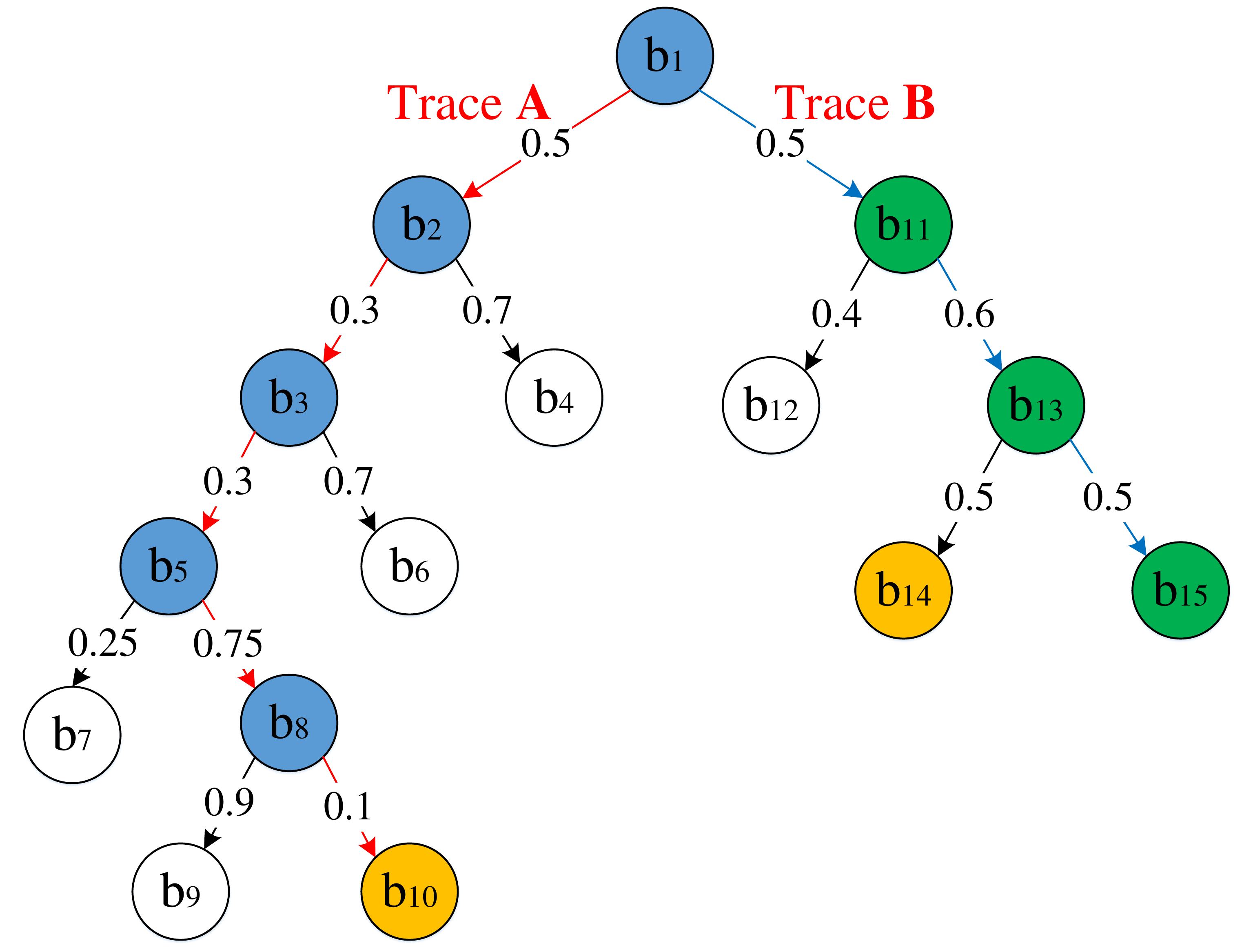}
	\caption{\label{fig:3}  The calculating method of path probability. }

\end{figure}

Since the fuzzer utilizes a random mutation strategy, the process of fuzzing can be viewed as a form of random sampling. As the number of samples increases in random sampling, the statistical probability will gradually approach the theoretical probability. Therefore, it is reasonable to expect that the statistical branch probabilities and path probabilities will converge toward their theoretical values with the increasing number of mutations in fuzzing.
The probability $P(br_j$) for all branches will be updated every minute. We have carefully selected this update interval to strike a balance between computational overhead and accuracy. A very short interval would lead to a high overhead in updating  $P(br_j$)  for all branches, whereas an overly long interval would result in inaccuracies when calculating $P(path_m$)  and the probability-based distance of each seed. Following thorough evaluations, we have concluded that updating $P(br_j$)  for all branches every minute is the optimal choice to maintain this balance.

As shown in Figure \ref{fig:3}, we use a simple artifactual C program as an example to illustrate how we calculate branch probability and path probability. In Figure \ref{fig:3}, each node (e.g., $b_1$) represents a basic block, and each edge (e.g., $<$$b_1, b_2$$>$) represents a basic block transition (i.e., branch). The path towards $b_5$ is a sequence of basic blocks starting from the entry basic block $b_1$, represented as $<$$b_1, b_2, b_3, b_5$$>$. The digitals above the line (e.g., $0.5$ above branch $<$$b_1, b_2$$>$) represent the branch probabilities evaluated based on statistical probabilities. For example, if the branch hits of $<$$b_1, b_2$$>$ and $<$$b_1, b_{11}$$>$ are both 50,000, the branch probabilities of the two branches are both $0.5$. After obtaining the branch probabilities, we can calculate the path probabilities of different basic blocks. For instance, the statistical path probability of $b_5$ is calculated as 0.5$\times$0.3$\times$0.3 = 0.045. Given the high throughput of fuzzing and the large number of random samples generated over time, it is reasonable and accurate to calculate branch and path probabilities based on statistical branch hits.
			
\subsubsection{Calculation of probability-based distance}
After obtaining the path probabilities of basic blocks, we combine them with BB distances to calculate the probability-based distances. The design is based on two considerations: (1) the probability-based distance is positively correlated with distance and negatively correlated with path probability, (2) introducing a factor of path probability
to establish both an upper bound and a lower bound for the distance. This allows us to control the impact of path probability and amplify the effect of changes in path probability, as the probability-based distance will exponentially change with path probability.

\begin{equation}\label{eq4}
	{d_p}(m,{T_b}) = {d_b}(m,{T_b}) \cdot {{\mathop{\rm c}\nolimits} ^{-P(path_m)}}
\end{equation}	    
Where $d_p(m,T_b)$ denotes the probability-based distance, $T_b$ denotes the target basic block, $d_b(m, T_b)$ denotes the BB distance. {$c^{(- P(path_m))}$  denotes the factor of path probability to establish both an upper bound ($d_b(m, T_b)$) and a lower bound ($\frac{c}{d_b(m, T_b)} $) for the distance. $P(path_m)$ denotes the path probability of $m$, and \textbf{c} is a constant which is greater than 1.

\subsubsection{Calculation of seed distance}
Up to now, almost all existing DGF techniques have used the arithmetic mean of all or part of the BB distances to calculate the seed distance. However, we have observed that in the seed's path, there are some basic blocks that are very close to or have already reached the target sites. These basic blocks contribute more to reaching and testing the target sites and we name them as \textbf{\textit{critical basic blocks}}. Since the arithmetic mean mainly reflects the general level of the population or the central tendency of the distribution \cite{64}, it cannot reflect the existence of critical basic blocks in the path.
\textbf{Therefore, we use the geometric mean, which is more sensitive to the minimum value, to calculate the seed distance.} For example, in Figure \ref{fig:3}, Trace A ($<$$b_1$, $b_2$, $b_3$, $b_5$, $b_8$, $b_{10}$$>$) is covered by Seed A, Trace B ($<$$b_1$, $b_{11}$, $b_{13}$, $b_{15}$$>$) is covered by Seed B, and $b_{10}$ and $b_{14}$ are the target basic blocks. Based on the arithmetic mean, the seed distance of Seed A ($\mathrm{(0+1+2+3+4+3)/6=2.16}$) is greater than the seed distance of Seed B ($\mathrm{(1+2+3)/3=2}$), resulting in Seed B being preferred. However, since Seed A can reach the target basic block, prioritizing Seed A is more reasonable. Given using the geometric mean, the seed distance of Seed A ($\sqrt[6]{0\times1\times2\times3\times4\times3}\ =0$) will be smaller than the seed distance of Seed B ($\sqrt[3]{1 \times 2 \times 3} = 1.81$). We can see that, in this example, the geometric mean more accurately reflects the presence of critical basic blocks than the arithmetic mean. For this reason, we use the geometric mean to calculate the seed distance.

\begin{equation}
	\label{eq5}
	\begin{small}
		{d_s}(s,{T_b}) = \left\{ \begin{array}{c}
			0,\;\;if\;{d_p}({m_i},{T_b})==0\; \wedge \;{m_i}\in{\xi _b}(s) \\
			\sqrt[{|{\xi _b}(s)|}]{{\prod\nolimits_{m \in {\xi _b}(s)} {{d_p}(m,{T_b})} }}, \ else
		\end{array} \right.
	\end{small}
\end{equation}

Where $s$ denotes the seed, ${\xi _b}(s)$ denotes the set of basic blocks in the execution path of the seed, and $|{\xi _b}(s)|$ denotes the number of basic blocks in ${\xi _b}(s)$. If there is a basic block $m_i$ whose BB distance is 0, it means that the current seed has hit a target basic block. In this case, we consider its seed distance as 0.

\textbf{Calculation Simplification}. To avoid the high overhead caused by the Product and Sqrt operations, we combine Equation (\ref{eq4}) and Equation (\ref{eq5}) to simplify the calculation of the geometric mean.
\begin{equation}
	\label{eq6}
\begin{small}
{d_s}{\rm{ = }}\left\{ \begin{array}{l}
			0,\;\;if\;{d_p}({m_i},{T_b})==0\; \wedge \;{m_i}\in{\xi _b}(s) \\
		{\rm{exp}}\left\{ {\frac{{\sum\nolimits_{m \in {\xi _b}(s)} {(\log ({d_b}(m,{T_b}) \ - \ P(path_m)} )}}{{|{\xi _b}(s)|}}} \right\}, \ else
						
		\end{array} \right.
	\end{small}
\end{equation}
We mainly simplify the second term in Equation (\ref{eq5}) based on Equation (\ref{eq4}). First, we take the logarithm of both sides of the equation in Equation (\ref{eq4}) to convert the Product and Sqrt operations into Summation operations, which yields ${\sum\nolimits_{m \in {\xi b}(s)} {\log ({d_b}(m,{T_b}) \cdot {{\mathop{\rm c}\nolimits} ^{-P(path_m)}})} }$. Then, to optimize the computation process, we set the constant $c$ in Equation (\ref{eq3}) to \textbf{e}, so that we can convert the multiplication operation in ${\log ({d_b}(m,{T_b}) \cdot {{\mathop{\rm c}\nolimits} ^{-P(path_m)}})}$ into an addition operation, which yields $\sum\nolimits{m \in {\xi _b}(s)} {(\log ({d_b}(m,{T_b}) \ - \ P(path_m)} )$.  Finally, we exponentiate the equation variables and obtain the optimized formula shown in Equation (\ref{eq6}). Through this simplification method, the forking process only needs to perform addition operations, greatly reducing the computation overhead.

\subsection{Power Schedule Optimization}
\label{section:dmab}

Most of the existing DGF techniques use seed distance as the fitness metric, such as AFLGo and WindRanger, which explicitly divides the fuzzing process into the \textit{exploration} phase and \textit{exploitation} phase. The exploration phase is designed to uncover as many paths as possible (like many coverage-guided fuzzers), and DGF in this phase favors seeds that expose new paths and prioritizes them. Then, based on the known paths, the exploitation phase is invoked to drive the engine toward the target code areas. In this phase, DGF prioritizes seeds that are closer to the targets and assigns more energy to them. 
However, the constraints on the closer seeds' paths may be difficult for DGF to satisfy, leading to a failure in generating new seeds within the limited time budget. Therefore, it is challenging to assign reasonable energy for both phases. With the probability-based distance, we design a Directed Multi-Armed Bandit (i.e., DMAB) model to optimize the power schedule, which can implicitly coordinate the exploration and exploitation in DGF. 

The Multi-Armed Bandit (i.e., MAB) problem results from the slot machine with multiple arms. The player plays one of the arms and obtains a reward. The player’s main goal is maximizing the rewards in finite trials.
Recent works, such as EcoFuzz \cite{36}, have applied the Adversarial MAB model to improve CGF, where the arms represent the seeds, and the reward represents the probability of uncovering new paths by mutating a seed.  
Based on the Adversarial MAB, EcoFuzz assumes that the probability of uncovering new paths decreases as the coverage of CGF increases.  
However, different from CGF, HyperGo measures the difficulty of covering new branches based on the branch probability of unexplored branches. Since the branch probabilities are only related to the complexity of condition constraints, which is a fixed value, we model the process of HyperGo covering new branches as a \textit{Stochastic Multi-Armed Bandit} problem.

In the stochastic MAB model, there are N fixed parallel arms, and at each time step \textit{t}, one arm indexed as $i$, ($i\in K={1,2,...., N}$) is selected to play. After playing arm \textit{i}, the player receives a reward, and the rewards of each slot machine may follow a fixed probability distribution. 
Using the greedy algorithm, we tend to select the arm with the highest reward. However, to obtain the global optimum, we need to explore different arms to evaluate their reward probability and select the arm with the highest reward expectation. In the DMAB model, we map the elements in DGF to rewards and reward probabilities and combine them into reward expectations. Compared to traditional DGF methods, DMAB eliminates the need to explicitly differentiate  exploration and exploitation. The assignment of seed energy is determined based on the difficulty of reaching the target sites for each seed, which implicitly incorporates the exploration and exploitation requirement.

\subsubsection{Elements in DMAB model}
We map the elements in DGF to the DMAB model as follows. 

\textbf{Reward}. We consider mutating seed $s$ as playing the slot machine. After mutating the seed \textit{s}, a new input is generated and a reward is obtained. The value of the reward depends on whether the new input covers a new branch. There are two possible values for the reward: 0 or a value \textit{r($d_s$)} which is negatively related to the seed distance, represented as $d_s$. The reward of 0 indicates that the new input  cannot cover a new branch while the reward of \textit{r($d_s$)} indicates that the new input covers a new branch. Based on a large number of tests, we find that new seeds generated from mutating the seed with shorter distance are more likely to have shorter distances. Thus, we believe that  the reward is negatively correlated with the seed distance of seed $s$. \textbf{To exploit the seeds with shorter distances, we prefer the seeds with higher rewards and assign the seeds more energy}.

\textbf{Reward probability}. To estimate the reward expectation, we need to evaluate the probability distribution of the reward. Whether DGF can cover new branches is related to whether the input can satisfy the constraint conditions of unexplored branches in the path. Therefore, we use the average branch probability of all unexplored branches in the path of seed \textit{s} to evaluate the probability of the fuzzer covering an unexplored branch through mutating seed $s$, represented as reward probability. \textbf{To explore more paths, we prefer the seeds that have higher probabilities of covering new branches and assign the seeds more energy.}

\textbf{Reward expectation}. We combine the reward and reward probability to evaluate the reward expectation. The design of the reward expectation is:

\vspace{-0.1cm}
\begin{equation}
	\label{eq7}
	E(r) =  {1 \over {{d_s}(s,{T_b})}} \cdot {{\sum\nolimits_{br \in \Theta (s)} {P(br)} } \over {|\Theta (s)|}}
\end{equation}
Where ${d_s}(s,{T_b})$ denotes the seed distance of the seed $s$,  $\Theta (s)$ denotes the set of unexplored branches in the path of $s$, $P(br)$ denotes the branch probability. The first term in Equation \ref{eq7} represents the reward, which is negatively correlated with ${d_s}(s,{T_b})$. The second term represents the reward probability, which is equal to the average branch probability of all unexplored branches. 

Based on Equation (\ref{eq7}), the DMAB model can dynamically coordinate the exploitation and exploration. As the fuzzing campaign launches, all seeds have not been sufficiently mutated and they have similar reward probabilities. Therefore, the DMAB model will exploit closer seeds with higher reward expectations. As fuzzing progresses, the closer seeds are sufficiently fuzzed and their reward probabilities will decrease as the number of mutations increases. According to Equation (\ref{eq7}), the reward expectation of these closer seeds will gradually become lower than that of the seeds with higher reward probability. As a result, the DMAB model will assign more energy to the seeds with higher reward probabilities to explore more new paths, implicitly switching to exploration.

\subsubsection{Design of power schedule}
After obtaining the reward expectations of all seeds, we optimize the power schedule based on the reward expectations to coordinate exploration and exploitation in DGF. We design the power schedule for two objectives. Firstly, the fuzzer should assign more energy to seeds with higher reward expectations, and less energy to seeds with lower reward expectations. Secondly, the energy of seeds can be adaptively adjusted with the progress of fuzzing. Based on these two objectives, we redesign the power schedule in HyperGo.

Firstly, we normalize the reward expectations of all seeds.
\begin{equation}
	\label{eq8}
	\tilde E(r) = {{E(r) - \min E} \over {\max E - \min E}}
\end{equation}
Where $\min E$ denotes the minimum reward expectation among all seeds, and $\max E$ denotes the maximum reward expectation among all seeds. Then, we integrate the normalized reward expectations with AFL's power schedule to form an optimized power schedule.

\begin{equation}
	\label{eq9}
	P(s,{T_b}) = \left\{ \begin{array}{l}
		{P_{afl}}(s) \cdot {2^{10 \cdot {\widetilde {E}(r)} - 5}}\;\;,if\;\;\Theta (s) \ne \emptyset \\
		\;\;\;\;\;\;\;\;\;\;\;\frac{{{P_{afl}(s)}}}{{32}} \;\;\;\;\; \;\;\;\;,if\;\;\Theta (s){\rm{ =  = }}\emptyset 
	\end{array} \right.
\end{equation}
Where $P_{afl}(s)$ denotes seed energy assigned by AFL's power schedule, $P(s,{T_b})$ denotes the finally assigned seed energy after optimization. AFL's power schedule assigns basic energy to the seed based on the seed's characteristics, such as the seed's execution speed and the size of its bitmap. By taking into account both the seed's characteristics and the reward expectation, we integrate AFL's power schedule and reward expectations to design the optimized power schedule. Moreover, to prevent the overuse of seeds and neglect of seeds that may contribute more to reaching target sites, we design the term ${2^{10 \cdot {\widetilde {E}(r)} - 5}}$ to control the adjustment of AFL's assigned energy within the range of [1/32, 32]. 

As fuzzing progresses, the number of unexplored branches and branch probabilities will change. This means that the reward expectations of all seeds will also change constantly. This allows HyperGo to dynamically assign seed energy and balance the trade-off between exploration and exploitation in a more accurate way. In Section \ref{section:effective}, we demonstrate the effectiveness of the DMAB model in steering DGF to reach the target sites.

\subsection{Optimized Symbolic Execution Complementary Scheme}
\label{section:osec}
To address the issues of directed hybrid fuzzing mentioned in Section \ref{background}, we design an Optimized Symbolic Execution Complementary (i.e., OSEC) scheme. In the OSEC scheme, we take three measures: pruning the unreachable branches, pruning the unsolvable branches, and dynamically prioritizing the symbolic execution of seeds. Algorithm \ref{alg:osec} represents the workflow of the OSEC scheme.  
\begin{algorithm}[h]
\vspace{-0.1cm}
	\footnotesize
	\caption{The Workflow of OSEC Scheme}
	\label{alg:osec}
	\begin{algorithmic}[1] 
		\REQUIRE $\Omega_s$, ${D}$
		\ENSURE $\psi_s $
		\WHILE {$s \in \Omega_s$}
		  \STATE $d_s, \bar P(\Theta (s)) \leftarrow D (s)$ 
		  \STATE Score(s) = \textbf{Cal\_Sco}($d_s$, $\bar P(\Theta (s))$, $\overline{SA} (\Theta (s))$)
		  \STATE \textbf{Sort}($\Omega_s$)
		\ENDWHILE
		\WHILE {true}
		    \STATE s $\leftarrow$ \textbf{Top\_Rank}($\Omega_s$)
		    \STATE  $\Theta (s)$ $\leftarrow$ \textbf{Sym\_Exe}(s) 
	        \WHILE {$br \in \Theta (s)$}
	          \IF {$br$ is unreachable $\textbf{or}$ $br$ is unsolvable} 
	            \STATE $\textbf{delete}$ $br$ from $\Theta (s)$
	          \ELSE
	             \STATE $new\_seed$ $\leftarrow$ \textbf{Constraint\_Solve}($\tau_{br}$)
	             \STATE $\psi_s $ = $\psi_s $ $ \cup $ $\{new\_seed\}$
	          \ENDIF
	        \ENDWHILE
	        \STATE Score(s) = \textbf{Cal\_Sco}($d_s$, $\bar P(\Theta (s))$, $\overline{SA} (\Theta (s))$)
	        \STATE \textbf{Sort}($\Omega_s$)
	    \ENDWHILE
	\end{algorithmic}
\vspace{-0.1cm} 
\end{algorithm}

In Algorithm 1, $\Omega_s$ represents the set of all seeds provided by DGF. $D$ represents a dictionary containing (\textit{seed index}, \textit{seed distance}, \textit{average branch probability}) triplets. Based on the seed's index, we can obtain the seed distance and the average branch probability of all unexplored branches in the seed's path (represented as $\bar P(\Theta (s))$). 
$\psi_s$ represents the set of all new seeds generated by the symbolic executor, $\overline{SA} (\Theta (s))$ represents the average number of solving attempts for unexplored branches, and $\tau_{br}$ represents the set of the unexplored branch's path constraints. Before the symbolic execution of seeds, the seed's $\bar P(\Theta (s))$ is calculated by DGF based on Equation (\ref{eq7}) and the $\overline{SA} (\Theta (s))$ is initialized to 1.

Firstly, the OSEC calculates the priority scores based on the seed distance, the average branch probability, and the average solving attempts. The priority scores are used to sort the seeds in descending order (Lines 2-5). Then, the OSEC continuously selects the seed with the highest priority score from $\Omega_s$, tracks the execution path of the seed, and identifies all unexplored branches to form the set $\Theta (s)$ (Lines 7-8). Next, the OSEC determines whether the unexplored branches are unsolvable or unreachable. If so, the OSEC prunes these branches. If not, the OSEC invokes the constraint-solver to solve the path constraints of these branches to generate new seeds. The new seeds are added to the set of new seeds $\psi_s$ (Lines 9-16). After the symbolic execution of the seed, the value of $d_s$, $\bar P_{br}(\Theta (s))$, and $\overline{SA} (\Theta (s))$ will all change. Therefore, we recalculate the priority score of the seed and re-sort all seeds (Lines 17-18). In the following sections, we will introduce in detail how to prune branches and calculate seed priority scores.

\subsubsection{Pruning the unreachable branches}
\label{section:pub}
\vspace{-0.2cm}
To avoid solving path constraints of the branches that do not contribute to reaching the target sites (i.e., unreachable branches), we need to prune the unreachable branches. We determine whether an unexplored branch is an unreachable branch based on the reachability of its destination basic block. We assume that if the destination basic block of a branch is unreachable, all successor basic blocks of that destination basic block are also unreachable. Thus, we consider this branch as an unreachable branch and give up solving its path constraints.

To determine whether an unexplored branch, represented as $<$$m_s, m_d$$>$, is unreachable, the OSEC loads the mappings of basic block addresses and BB distances ($<$$BB\_add, BB\_dis$$>$) provided by the static analyzer to obtain the BB distances of all basic blocks. Then, the OSEC checks the BB distance of $m_d$ (whether $d_b(m_d,T_b) \ge 0 $) to determine the reachability of branch $<$$m_s, m_d$$>$. If branch $<$$m_s, m_d$$>$ is unreachable, the symbolic executor will abandon solving this branch and prune it from the set of unexplored branches. 
\vspace{-0.2cm}
\subsubsection{Pruning the unsolvable branches}
To  alleviate the issue of solving path constraints for the branches that cannot be solved (i.e., unsolvable branches) within a limited time budget, we need to prune the unsolvable branches. When solving the path constraints of a branch, we make two basic assumptions. (1) The time budget for solving the path constraints of each branch has a lower bound (e.g., 5s) and an upper bound (e.g., 15min). The time budget increases with the number of solving attempts, up to the upper limit. (2) If the symbolic executor fails to solve a branch due to the complexity of the branch's path constraints,
all subsequent branches of that unsolvable branch are also unsolvable since they have more path constraints than the unsolvable branch.

Based on these two assumptions, we can prune the unsolvable branches.
Firstly, we dynamically adjust the time budget within the ranger of $[lower\_bound, upper\_bound]$ according to the solving attempts. Based on empirical experience, we increase the time budget by 1 minute after each attempt. If the branch cannot be solved within the upper limit time budget, we consider it as an unsolvable branch and give up solving it. Secondly, during the symbolic execution of a seed, we would record the solving results (e.g., success or failure) of the unexplored branches that have been identified. Then, before solving each branch, we check whether its predecessor branch is unsolvable. If unsolvable, based on the second assumption, the path constraints of this branch are too complicated for the symbolic executor to solve within the limited time budget, and thus the branch is regarded as an unsolvable branch. Through these two steps, we prune the unsolvable branches to alleviate the pressure of constraint solving.

\subsubsection{Prioritizing the symbolic execution of seeds}
To achieve a complementary integration of DGF and SE, we need to dynamically adjust the order of seeds for symbolic execution. Due to its high fuzzing throughput, DGF can quickly cover paths with easy-to-satisfy path constraints. In contrast, SE has the excellent constraint-solving ability to generate inputs that satisfy complex path constraints. To fully leverage the respective strengths of DGF and SE, we want (1) both SE and DGF to prioritize seeds with shorter seed distances, (2) SE to prioritize solving unexplored branches that are difficult for DGF to cover, and (3) SE to prioritize seeds with fewer solving attempts and lower time budgets. We calculate the priority scores of different seeds based on these three considerations and adjust the order of seeds for symbolic execution accordingly.

Firstly, similar to the method of evaluating reward probability in Section \ref{section:dmab}, we use the branch probability of unexplored branches to evaluate the difficulty of the fuzzer covering the unexplored branches through mutating seed $s$.

\begin{equation}
	\label{eq10}
	EDF(s) = {{\sum\nolimits_{br \in \Theta (s)} {P(br)} } \over {|\Theta (s)|}}
\end{equation}
Where $EDF(s)$ denotes the estimated difficulty. Similarly, we use the average branch probability of all unexplored branches in the path of seed $s$ as the estimated difficulty. Moreover, based on the branch probability calculation method in Section \ref{section:distance}, the branch probabilities of unexplored branches are all greater than 0.

Then, we evaluate the solving difficulty of branches based on the number of solving attempts. That is, more solving attempts indicate that the symbolic executor has more difficulty in solving the path constraints of the branch. We use the average solving attempts of all unexplored branches in the path of seed $s$ to evaluate the solving difficulty of the seed's branches.

\begin{equation}
	\label{eq11}
	EDS(s) = {{\sum\nolimits_{br \in \Theta (s)} {SA(br)} } \over {|\Theta (s)|}}
\end{equation}
Where EDS(s) denotes the estimated difficulty, $SA(br)$ denotes the number of solving attempts of the unexplored branch.

Based on $EDF(s)$, $EDS(s)$, and $d_s(s, T_b)$, we score the seeds to determine their order for symbolic execution. To ensure that the three indicators have the same weight in affecting the priority score of seeds, we use a normalization method as shown in Equation (\ref{eq8}) to calculate their normalized value. Then, we can calculate the priority score of different seeds. 
\begin{equation}
\label{eq12}
Score(s) = {{\widetilde {EDF}(s)} \over {\widetilde {EDS}(s) \cdot \widetilde {d_s}(s,{T_b})}}  
\end{equation}
Where $\widetilde {EDF}(s)$, $\widetilde {EDS}(s)$ and $\widetilde {d}(s,{T_b})$ are the normalized value. After obtaining the priority scores of all seeds, the OSEC scheme will prioritize the seeds with higher scores for symbolic execution. Moreover, since the three factors used to calculate the priority scores, ${EDF}(s)$, ${EDS}(s)$, and ${d}(s,{T_b})$, are changing dynamically, the OSEC scheme will adaptively adjust the order of symbolic execution of the seeds. By this method, the OSEC scheme dynamically prioritizes the optimal seeds whose unexplored branches are hard for the fuzzer to penetrate through, are more likely to be solved by the symbolic executor, and are closer to targets.  

During the symbolic execution, the symbolic executor will constantly attempt to solve the unexplored branches in the seeds' paths and generate new interesting seeds for DGF.

\section{Implementation}

The implementation of HyperGo mainly consists of three components: a static analyzer, a fuzzer, and a symbolic executor. For the static analyzer, we leverage the static analysis framework LLVM 11.0 and Clang 11.0 and use the LLVM IR to instrument the program. The fuzzer is built on AFL 2.52b, and the symbolic executor is built on Symcc. The implementation part of HyperGo is implemented with about 2000 lines of C/C++ and RUST code. 
HyperGo is publicly available on our website (\url{https://gitee.com/paynelin/hypergo}).

\section{Evaluation}
To evaluate the effectiveness of HyperGo, we conducted experiments aiming to answer five research questions:

\noindent \textbf{RQ1:} What about the performance of HyperGo in terms of reaching the target sites?

\noindent \textbf{RQ2:} What about the performance of HyperGo in terms of exposing the vulnerabilities in the target sites?

\noindent \textbf{RQ3:} How probability-based distance, the DMAB model, and the OSEC scheme take effect in the overall performance of HyperGo?


\noindent \textbf{RQ4:} Is the probability-based fitness metric effective in finding better seeds for directed fuzzing?

\noindent \textbf{RQ5:} What about the performance of HyperGo in terms of discovering new vulnerabilities?
\graphicspath{{images/} }

\subsection{Evaluation Setup}

\textbf{Evaluation Criteria.} We mainly use two types of criteria to evaluate the performance of different fuzzing techniques.

(1)Time-to-Reach (TTR) is used to evaluate the time spent on generating the first input which can reach the specific target site. 

(2)Time-to-Expose (TTE) is used to evaluate the time spent on exposing the (known or undisclosed) vulnerabilities in the target sites. When a crash is observed at the target site, it indicates that the fuzzer has successfully exposed the vulnerability.

\textbf{Evaluation Benchmarks.} We selected two datasets and 7 real-world programs with potential vulnerabilities.

(1) UniBench \cite{42} provides real-world programs of different types and the corresponding seed corpus. The state-of-the-art fuzzing techniques, such as WindRanger, have used the UniBench as the benchmark for testing. To answer RQ1, RQ3, RQ4, and RQ5, we tested the 20 programs from UniBench and used AFL++ \cite{AFL++} to select target sites from each program by conducting preliminary experiments. We first ran AFL++ for 48 hours and collected all the seeds generated by AFL++. Then, we use afl-cov to re-run these seeds, so that we can obtain the code locations covered and the time when they are covered, represented as pairs like (line, time). Finally, among the locations that are reached using from 1 hour to 48 hours (i.e., more than 1 hour), we randomly selected 4 code locations as the targets.

(2) AFLGo testsuite \cite{43} was proposed in AFLGo's paper and website to evaluate the directness of DGF, and it had been used as a benchmark by many state-of-the-art directed fuzzers (e.g., Hawkeye and WindRanger). To answer RQ2, we selected it as the benchmark.

(3) Additional real-world programs. In addition to UniBench, we add another 9 real-world programs to construct a new testbench (all the programs are listed in Table \ref{table:10} in the Appendix). 
To answer RQ4, we used sanitizers (e.g., UBSAN \cite{53}, ASAN\cite{54}) to label the target sites for the testbench and tested them with HyperGo.

\textbf{Baselines.} In our evaluation, we compared HyperGo with the state-of-the-art directed greybox fuzzers that are publicly available by the time of writing this paper, including WindRanger, BEACON, ParmeSan, and AFLGo. To conduct the incremental experiments, we combined AFLGo with SymCC to form a new directed hybrid fuzzer, which is called AFLGoSy, as the baseline.

\textbf{Experiment Settings.} We conducted the experiments on the machine equipped with Intel(R) Xeon(R) Gold 6133 CPU @ 2.50GHz with 80 cores and used Ubuntu 20.04 LTS as the operating system. All the experiments were repeated \textbf{5} times within a time budget of \textbf{24 hours}. When testing the programs we used the seeds in the BenchMarks' recommended seed corpus as initial seeds. 
Given that HyperGo requires two CPU cores to simultaneously launch both fuzzing and symbolic execution instances, the compared fuzzers also employed parallel fuzzing by launching two fuzzing instances (one acting as the master instance and the other as the slave instance).
For experimental results analysis, we utilize the Mann-Whitney U test (p-value) to measure the statistical significance and the Vargha-Delaney statistic ($\hat{A}_{12}$) \cite{39} to measure the probability of one technique performing better than another.

\subsection{Reaching Target Sites} 
\label{reach}

\begin{figure}[t]
	\centering
	\setlength{\abovecaptionskip}{0.1cm}
	\includegraphics[width=\columnwidth]{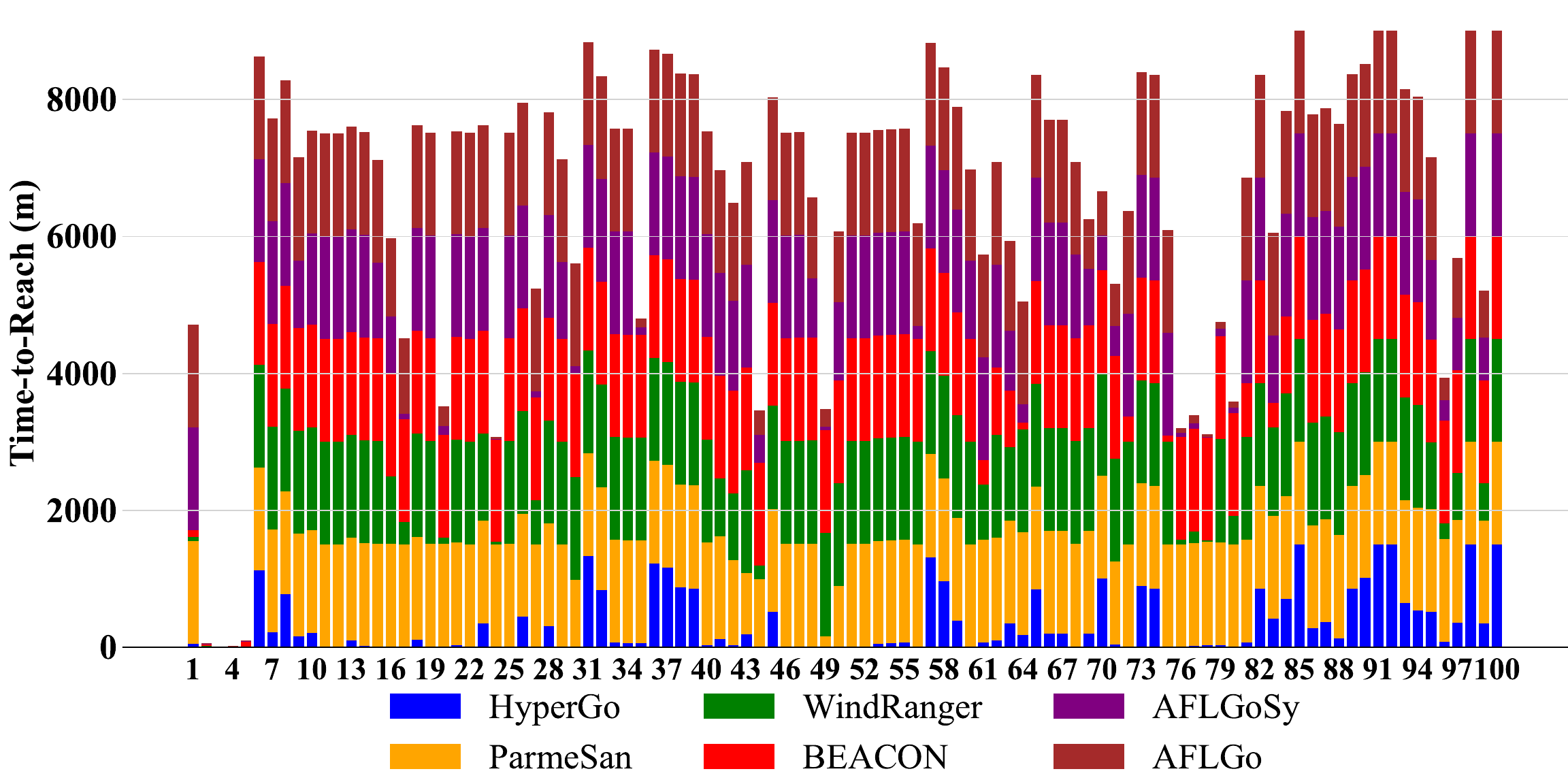} 
	\caption{\label{fig:ttr}  TTR of AFLGo, AFLGoSy, BEACON, WindRanger, ParmeSan, and HyperGo on the UniBench.} 
\end{figure}

\noindent To answer RQ1, we tested programs from UniBench, with a total of 100 target sites, and evaluated the TTR of different fuzzers. We set the timeout threshold as 24 hours. 
The detailed results of TTR are listed in Table \ref{table:unibench} in the Appendix.
In Table \ref{table:unibench}, the entry ``N/A" indicates that the fuzzer failed to compile the program due to code issues, while ``T.O." indicates that the fuzzer couldn't reach the target site within the allocated 24-hour time budget. For WindRanger, some entries are marked as ``N/A" due to encountering segmentation fault errors or being unable to obtain distance information during program testing. As for BEACON and ParmeSan, most entries showing ``N/A" might be because it is incompatible with UniBench. 
For ``N/A" entries, we did not use them to calculate the speedups and p-values. 
As for the ``T.O." entries, we believe that these fuzzers might still reach the targets in subsequent fuzzing processes. Therefore, we opted for a slightly larger value of 1500 minutes to calculate speedups and p-values.

According to the results of TTR, HyperGo can reach the most (95/100) target sites compared to AFLGo (28/100), AFLGoSy (38/100), BEACON (14/100), WindRanger (25/100), and ParmeSan (11/100) within the time budget. Moreover, on most of the target sites (89/100), HyperGo outperforms all other fuzzers and achieves the shortest TTRs. In terms of mean TTR of reaching the target sites, HyperGo demonstrates 37.75$\times$, 29.11$\times$, 23.34$\times$, 95.61$\times$ and 143.22$\times$ speedup compared to AFLGo, AFLGoSy, BEACON, WindRanger, and ParmeSan, respectively.
We conducted both the Mann-Whitney U test (p-value) and the Vargha-Delaney test ($\hat{A}_{12}$), all the p-values are less than 0.01, and the mean $\hat{A}_{12}$ against AFLGo, AFLGoSy, BEACON, WindRanger, and ParmeSan are 0.88, 0.85, 0.92, 0.86, and 0.91, respectively. Based on the above analysis, we can conclude that \textbf{HyperGo can reach the target sites faster than baseline fuzzers}.

To reflect the results in a straight way, we use bar charts to visualize the results. In Fig. \ref{fig:ttr}, the x-axis represents the target site ID (1-100), the y-axis represents the total TTR of all fuzzers in minutes, 
and a shorter bar indicates a shorter TTR. Since some fuzzers cannot compile some programs or reach the target sites within the 24-hour time budget, resulting no TTR. To distinguish these cases, the TTR of such a case is represented as 1500 minutes in Fig. \ref{fig:ttr}. From the figure, we can clearly see that the blue bars are much shorter than the other bars, which means that HyperGo can reach most of the target sites faster than the baseline fuzzers. 

\subsection{Exposing vulnerabilities} 
\label{expose}
\begin{table}[t]
	\vspace{-0.1cm}	
	\footnotesize
	\centering
	\setlength{\abovecaptionskip}{0.1cm} 
	\setlength{\belowcaptionskip}{0cm}
	\setlength{\tabcolsep}{2pt}
	\caption{The results of TTE on AFLGo testsuite}
	\label{table:6}
	
	\begin{tabular}{cccccccccc}
		\toprule[1.5pt]
		\textbf{Prog.}&   \textbf{CVE-ID}&{AFLG} & {AFLS} & {BEAC} & {Wind} & {Parm} & {HyGo} \\
		\midrule
		\multirow{7}{*}{binutils$_{2.26}$}   & 2016-4487 &2.33m & 2.42m & 0.63m & 1.21m & 0.95m  & 1.42m \\
		&2016-4488      &4.23m  & 3.60m & 32.1m & 3.32m &  2.62m  & 2.12m \\
		& 2016-4489    &3.36m & 4.11m & 2.98m & 5.88m  & 2.31m   & 1.89m \\
		& 2016-4490   &1.15m & 1.81m  & 2.35m & 2.63m  & 0.82m   & 1.68m  \\
		& 2016-4491    &448m  & 389m  &  258m & 298m  &  212m  & 69.3m  \\
		& 2016-4492      &10.8m & 13.2m & 43.6m  & 7.47m  &  4.33m  & 3.94m \\
		&   2016-6131     &348m & 236m & 292m &  318m & 244m  & 101m \\ \midrule
		\multirow{4}{*}{libming$_{4.48}$} & 2018-8807 &331m & 218m & 267m  & 171m & 301m   & 68.3m \\
		&2018-8962    &234m & 271m & 163m & 121m & 198m   & 43.7m \\    
		& 2018-11095     &T.O. & 914m & 252m & 1311m & T.O. & 118m \\
		& 2018-11225     &T.O. & T.O. & 438m & 996m  & T.O.  & 202m  \\ \midrule
		\multirow{3}{*}{LibPNG$_{1.5.1}$}  &2011-2501 &10.2m & 12.3m & N/A & 7.81m & 4.53m   & 2.16m \\
		&2011-3328     &69.1m & 54.3m & N/A & 49.3m & 193m   &  21.1m \\
		& 2015-8540  &0.88m & 1.19m & N/A & 0.96m  & 3.41m   & 2.65m \\ \midrule
		\multirow{4}{*}{xmllint$_{2.9.4}$}  & 2017-9047 &T.O. & T.O.  & T.O. & T.O. & T.O. & 983m \\
		& 2017-9048    &T.O. & T.O. & T.O. & T.O. &  T.O.  & T.O.  \\
		& 2017-9049    &T.O. & T.O. & T.O. & T.O. &  T.O.  & 635m \\
		& 2017-9050    &T.O. & T.O. & T.O. & T.O. & T.O.   &T.O. \\ \midrule
		\multirow{2}{*}{Lrzip$_{0.631}$}  & 2017-8846 &348m & 284m & 156m&  223m &  466m  & 69.4m \\
		&2018-11496    & 201m & 226m & 98.1m & 169m & 126m   & 33.9m \\
		
		\midrule
		\multicolumn{2}{c}{\fontsize{6}{5}\selectfont \textbf{speedup}}         &{\textbf{3.44$\times$}} & {\textbf{3.63$\times$}} & {\textbf{4.10$\times$}} & {\textbf{3.26$\times$}} & {\textbf{3.00$\times$}} & {\textbf{-}} \\ 
		\multicolumn{2}{c}{\fontsize{6}{5}\selectfont \textbf{mean $\hat{A}_{12}$}}      &{\textbf{0.84}} & {\textbf{0.82}} & {\textbf{0.79}} & {\textbf{0.76}} & {\textbf{0.80}} & {\textbf{-}} \\
		\multicolumn{2}{c}{\fontsize{6}{5}\selectfont \textbf{mean p-values}} & {\textbf{$0.009$}} & {\textbf{$0.013$}} & {\textbf{$0.006$}} & {\textbf{$0.026$}} & {\textbf{$0.008$}} & {\textbf{-}} \\  
		\bottomrule[1.5pt]
		
	\end{tabular}
	\label{table:tte}
	
\end{table}

\noindent To answer RQ2, following BEACON and WindRanger, we used the AFLGo testsuite and set the known vulnerabilities with CVE IDs in the programs as the target sites. The information on target sites and the TTE results
are presented in Table \ref{table:tte}. As Table \ref{table:tte} shows, among the 20 vulnerabilities, HyperGo exposed the most (18) compared to AFLGo (14), AFLGoSy (15), BEACON (13), 
WindRanger (16), and ParmeSan (14). Besides, on most of the target sites (15/20), HyperGo outperformed all the baseline fuzzers and achieved the shortest TTE. Among the 20 vulnerabilities, HyperGo costs longer time than baselines for three CVEs (2016-4487, 2016-4490, and 2015-8540). The three CVEs are swiftly discovered by all fuzzers within a few minutes of launching the fuzzing campaign.  During this initial period, the branch hits for all branches are insufficient to accurately assess branch probabilities and calculate path probabilities. Consequently, in some instances during these first few minutes, there is a possibility of the fuzzer and symbolic executor incorrectly prioritizing branches. However, as the fuzzing process continues and the branch hits increase, HyperGo would address this issue and perform better than the baseline fuzzers in exposing the deeper bugs.
With respect to the mean TTE of exposing vulnerabilities, HyperGo demonstrated 3.44$\times$, 3.63$\times$, 4.10$\times$, 3.26$\times$ and 3.00$\times$ speedup compared to AFLGo, AFLGoSy, BEACON, WindRanger, and ParmeSan, respectively. All p-values were less than 0.05, and the mean $\hat{A}_{12}$ against AFLGo, AFLGoSy, BEACON, WindRanger, and ParmeSan were 0.84, 0.82, 0.79, 0.76, and 0.80, respectively. Based on the above analysis, we can conclude that \textbf{HyperGo can expose known vulnerabilities faster than the baseline fuzzers}.

\begin{table*}[hbp]
	\tiny
	\centering
	\setlength{\abovecaptionskip}{0.1cm} 
	\setlength{\belowcaptionskip}{0cm}
	\setlength{\tabcolsep}{4pt}
	\caption{The TTR results on programs from UniBench}
	\label{table:unibench}
	
	\begin{tabular}{@{}ccccccccccccc@{}}
		
		\toprule[1.5pt]
		\textbf{No} & \textbf{Prog} & \textbf{Version} &\textbf{Target sites} & {AFLGo} & {AFLGoSy} & {BEACON} & {WindRanger} & {ParmeSan} &  {Only-PB} & PB+DMAB & {HyperGo} \\
		\midrule
		1  & \multirow{5}{*}{cflow} &    \multirow{5}{*}{1.6}  & parser.c:281 & T.O. & T.O. & 99.4m &61.1m & T.O. & 311m &224m& 51.8m \\
		2  &      &          & c.c:1783 & 12.8m & 9.43m & 22.1m & 6.45m & 10.1m   & 8.41m & 7.09m& 8.82m \\
		3  &      &          & parser.c:105 & 0.82m & 1.21m & 13.5m & 0.93m & 0.44m & 2.37m & 2.88m  & 1.68m \\
		4  &      &          &  parser.c:1223 &  1.22m & 1.66m & 0.83m & 2.44m & 6.23m  &  2.44m & 7.21m &  10.8m\\
		5  &      &          & parser.c:108 & 12.8m & 8.63m  & 68.1m & 8.32m & 6.76m  & 4.33m &2.69m &  1.65m \\ \hline
		
		6  & \multirow{5}{*}{mp42aac} &    \multirow{5}{*}{Bento4 1.5.1-628}      &   Ap4AvccAtom.cpp:82  &  T.O. &  T.O. & N/A & T.O. &  N/A   &T.O.  & T.O.& 1124m \\
		7  &      &        &   Ap4TrunAtom.cpp:139 &  T.O.&  T.O. & N/A & T.O. &   N/A  &T.O.  & T.O.& 223m  \\
		8  &      &        &  Ap4SbgpAtom.cpp:81&  T.O. & T.O.  & N/A & T.O. &  N/A   &T.O.  & T.O.& 781m  \\
		9  &      &        &  Ap4TfdtAtom.cpp:71 &  T.O.&  985m & N/A & T.O. &   N/A &304m & 287m & 166m  \\
		10 &      &       & Ap4AtomFactory.cpp:490 &  T.O.&  1324m &  N/A& T.O. &  N/A   &1318m & 1311m  & 215m \\ \hline
		11 & \multirow{5}{*}{jhead} &  \multirow{5}{*}{3.00}    &    exif.c:1339& T.O.& T.O. & N/A & T.O. & T.O.   & T.O. & T.O.   &5.52m \\
		12 &      &      &    exif.c:1327  &  T.O.& T.O. &N/A   &T.O.  & T.O. &  T.O.   & T.O.    & 2.74m \\
		13 &      &       &   iptc.c:143  &  T.O.& T.O. &N/A   & T.O. & T.O.   &T.O. & T.O.   & 107m \\
		14 &      &       &    iptc.c:91 &  T.O.& T.O. &N/A  & T.O. &  T.O.  & T.O. &771m   &20.8m \\
		15 &      &       &    makernote.c:174 &  T.O.& 1102m &N/A  & T.O. &  T.O.  & 417m &  349m  & 11.3m \\ \hline
		16 & \multirow{5}{*}{mp3gain} &  \multirow{5}{*}{1.5.2}    &     layer3.c:1116 &  1142m & 841m & N/A & 984m & N/A   & 262m & 229m   &10.84m \\
		17 &      &      &     interface.c:690 &  1098m & 81.9m & N/A & 324m & N/A   &67.6m &61.1   & 9.02m \\
		18 &      &       &   mp3gain.c:602 &  T.O. & T.O. & N/A & T.O. & N/A   &T.O.  & 771m   & 119m \\
		19 &      &      &    interface.c:663 &  T.O. & T.O. &  N/A& T.O. & N/A   &T.O. & T.O.   & 12.8m \\
		20 &      &      &   apetag.c:341 &  290m & 132m & N/A & 91.2m  &  N/A &132m  & 67.4m & 11.8m \\ \hline
		21 & \multirow{5}{*}{lame} &     \multirow{5}{*}{3.99.5}   &    bitstream.c:823 &   T.O. & T.O. &N/A  &T.O.  & N/A   & T.O.  & T.O.   &36.8m \\
		22 &      &        &   lame.c:2148 &   T.O. & T.O. & N/A & T.O. &  N/A  &T.O.  &  T.O.   &  8.73m\\
		23 &      &        &  uantize\_pvt.c:441 &   T.O. & T.O. & N/A &  1269m  & N/A &T.O.  &  T.O.   & 354m \\
		24 &      &        &  VbrTag.c:778  & 26.5 m & 1.42m & N/A & 39.1m &  N/A  &1.40m & 1.41m   & 2.96m\\ 
		25 &      &         &   get\_audio.c:1605 &  T.O.& T.O. &  N/A& T.O. &N/A & T.O. & T.O.  &11.5m \\ \hline
		26 & \multirow{5}{*}{imginfo} &  \multirow{5}{*}{jasper 2.0.12}  &    jp2\_cod.c:841 &T.O.& T.O. &  N/A&   T.O. &  T.O.  & T.O. &   T.O.  &451m \\
		27 &      &         &   jpc\_dec.c:1393 &  T.O.& 89.1m  & N/A & 653m & T.O.   &39.0m & 33.1m  &  0.35m \\
		28 &      &         &  jp2\_cod.c:636 & T.O. & T.O. &  N/A & T.O. & T.O.   &T.O. & T.O.    & 314m \\
		29 &      &         & jas\_stream.c:823 & T.O. & 1123m &N/A   & T.O. &  T.O.  &1101m   & 1088m   & 0.71m\\
		30 &      &         &  jpc\_dec.c:1393& T.O. & 121m & N/A &  T.O.  & 984m   &26.8m &  23.0m    & 0.81m \\ \hline
		31 & \multirow{5}{*}{gdk-pixbuf-pixdata} & \multirow{5}{*}{gdk-pixbuf 2.31.1}  &    gdk-pixbuf-loader.c:387 & T.O. & T.O. & T.O. & T.O. & N/A & T.O. &  T.O. &1339m  \\
		32 &      &         &   io-qtif.c:511  &T.O.  & T.O. &  T.O. & T.O. & N/A &T.O. &  T.O. & 841m \\
		33 &      &         &  io-ani.c:403 &  T.O.&  T.O.&  T.O. & T.O. & N/A &  T.O. &  T.O. &72.3m \\
		34 &      &         & io-jpeg.c:691 &  T.O.&  T.O.&  T.O. & T.O. & N/A & T.O. &  T.O. &68.6m  \\
		35 &      &         &  io-tga.c:360 & 126m & 111m &  T.O. & T.O. &  N/A &  106m & 74.1m &60.7m \\ \hline
		36 & \multirow{5}{*}{jq} &   \multirow{5}{*}{1.5}  &    jv\_dtoa.c:3122 & T.O. &  T.O. & T.O. & N/A & T.O. &T.O. & 1241m &1223m   \\
		37 &      &         &   jv\_dtoa.c:2004 & T.O. &  T.O. & T.O. & N/A & T.O. & T.O. & T.O.  &1163m  \\
		38 &      &         & jv\_dtoa.c:2518 & T.O. &  T.O. & T.O. & N/A & T.O. & T.O. & T.O.  & 875m \\
		39 &      &         & jv\_unicode.c:42 & T.O. &  T.O. &T.O.  &N/A  & T.O.  & T.O. & T.O.  & 864m \\
		40 &      &          &  jv\_dtoa.c:3044  & T.O. &   T.O.& T.O. & N/A & T.O. & T.O. & 929m & 37.1m \\ \hline
		41 & \multirow{5}{*}{tcpdump} &   \multirow{5}{*}{4.8.1}  &     print-aodv.c:259 & T.O. &  T.O. & N/A & 843m  &T.O.  &T.O.  &T.O.  & 124m  \\
		42 &      &       &     print-ntp.c:412 & 1436m & 1311m & N/A & 974m &1239m   &801m &383m   & 33.7m  \\
		43 &      &        &     print-rsvp.c:1252  & T.O. &  T.O. & N/A & T.O.  &889m & T.O.  & T.O.  & 194m \\
		44 &      &         &     print-snmp.c:607 & 359m & 412m & N/A & 192m &992m &124m &992m & 11.2m  \\
		45 &      &         &     print-l2tp.c:606  & T.O. &  T.O. & N/A & T.O.  &T.O. &T.O.  & T.O.   &  526m \\ \hline
		46 & \multirow{5}{*}{tic} &   \multirow{5}{*}{ncurses 6.1}    &     captoinfo.c:189 & T.O. &  T.O. &N/A  & N/A & T.O. &T.O. & T.O. & 15.1m  \\
		47 &      &         &     alloc\_entry.c:141 & T.O. &  T.O. & N/A &  N/A&T.O.  &T.O.&T.O.  &  19.1m \\
		48 &      &         &     name\_match.c:111 & 1186m & 866m &N/A  &N/A  &T.O.  &623m  &86.6m &  19.5m \\
		49 &      &         &    comp\_scan.c:860  &  264m & 49.3m  & N/A & N/A &168m  &14.3m &7.24m &  1.10m \\
		50 &      &          &     entries.c:78 &  1038m & 1134m & N/A & N/A & 883m &824m & 336m &  19.9m \\ \hline
		51 & \multirow{5}{*}{flvmeta} &   \multirow{5}{*}{1.2.1}   &     json.c:1036 &T.O.  &  T.O. & N/A & T.O. & T.O. &T.O.  & T.O. & 10.1m   \\
		52 &      &        &     avc.c:1023  & T.O. &  T.O. &  N/A&  T.O.& T.O. &T.O. & T.O. &  12.9m  \\
		53 &      &         &     api.c:718 &T.O.  &  T.O. & N/A &  T.O.& T.O. &T.O. & T.O. &   50.1m \\
		54 &      &        &     flvmeta.c:1023 & T.O. &   T.O.& N/A &  T.O.&  T.O.& T.O. &  T.O.&  60.6m \\
		55 &      &         &     check.c:769 & T.O. &  T.O. &  N/A&  T.O.& T.O. &T.O.& T.O. &   74.7m \\ \hline
		56 & \multirow{5}{*}{tiffsplit} &   \multirow{5}{*}{libtiff 3.9.7}  &     tif\_ojpeg.c:1277 & T.O. &  188m & N/A & T.O. &  N/A&79.7m &60.4m &  7.41m   \\
		57 &      &          &     tif\_read.c:335 &T.O.  &  T.O. &N/A  & T.O. & N/A &T.O.   & T.O.  &  1321m  \\
		58 &      &         &     tif\_jbig.c:277 & T.O. &  T.O. & N/A & T.O.  & N/A & T.O. & T.O.   & 967m  \\
		59 &      &         &     tif\_dirread.c:1977 & T.O. & T.O. & N/A & T.O. & N/A &T.O. & T.O.  &  388m  \\
		60 &      &         &     tif\_strip.c:154 & 1328m &1139m  & N/A & T.O. &  N/A& 926m &  796m & 8.73m \\ \hline
		61 & \multirow{5}{*}{nm} &  \multirow{5}{*}{binutils-5279478}  &     tekhex.c:325 & T.O. &  T.O. & 364m  & 798m  & N/A &  T.O. & T.O. &78.4m  \\
		62 &      &         &     elf.c:8793 & T.O. &  T.O. & 986m & T.O. & N/A &T.O. & T.O. &   102m \\
		63 &      &         &     dwarf2.c:2378  & 1313m & 868m & 831m & 1065m &  N/A&553m & 512m & 357m  \\
		64 &      &          &     dwarf1.c:281 & T.O. &  268m & 98m & T.O.  &  N/A& 241m & 233m  & 187m \\
		65 &      &         &     elf-properties.c:51 & T.O. &  T.O. &T.O.  & T.O. &  N/A& T.O. &T.O.  & 853m \\ \hline
		66 & \multirow{5}{*}{pdftotext} &  \multirow{5}{*}{4.00}   &     XRef.cc:645  & T.O. &  T.O. &  N/A & T.O. &  N/A& T.O.  & T.O. &201m  \\
		67 &      &         &     Stream.cc:2658  & T.O. &  T.O. &  N/A & T.O. &  N/A&T.O.  & T.O. & 201m \\
		68 &      &         &     GfxFont.cc:1337 & 1345m & 1223m &  N/A & T.O. & N/A & 1208m & 579m  & 17.5m \\
		69 &      &         &     Stream.cc:1004 &  725m& 824m &  N/A & T.O. & N/A&   631m & 330m &207m \\
		70 &      &         &     GfxFont.cc:1643 &  637m&  514m &   N/A& T.O. &  N/A& 477m & 586m & 1004m \\ \hline
		71 & \multirow{5}{*}{sqlite3} &   \multirow{5}{*}{SQLite 3.8.9}   &     pager.c:5017 & 617m & 436m & N/A & N/A & 1214m & 196m & 174m &44.1m \\
		72 &      &         &     select.c:4301 & T.O. &  T.O. & 367m  & N/A & T.O. & T.O.  & T.O. & 2.55m \\
		73 &      &         &     func.c:1029 & T.O. &  T.O. & T.O. & N/A & T.O. &T.O. & T.O.  &  896m \\
		74 &      &         &     insert.c:1498 & T.O. &   T.O.& T.O. & N/A & T.O.  &T.O.& T.O. &  857m \\
		75 &      &         &     vdbe.c:1984  & T.O. &   T.O.& 89.6m &N/A  & T.O. &T.O.  & T.O. & 0.90m  \\ \hline
		76 & \multirow{5}{*}{exiv2} &   \multirow{5}{*}{0.26}    &     tiffcomposite.cpp:82 & 73.1m & 59.3m &N/A  & 68.1m & N/A&26.0m & 27.6m &  1.90m  \\
		77 &      &        &     XMPMeta-Parse.cpp:1037  & 126m &  78.1m& N/A & 168m & N/A &  72.9m  &60.5m  & 21.8m \\
		78 &      &        &     XMPMeta-Parse.cpp:847 & 37.5m & 13.4m &N/A  & 21.4m &N/A & 16.2m  & 14.3m  & 39.8m \\
		79 &      &        &     tiffvisitor.cpp:1044 & 102m&   111m &N/A& T.O. &  N/A& 109m & 93.8m &  39.5m \\
		80 &      &         &    XMPMeta-Parse.cpp:896 & 86.7m &  78.4m & N/A & 421m & N/A&   95.3m & 57.2m &  1.72m \\ \hline
		81 & \multirow{5}{*}{objdump} &  \multirow{5}{*}{binutils-2.28}  &     elf.c:9509 & T.O. &  T.O. &782m &T.O. &T.O. &T.O. &T.O. & 78.4m  \\
		82 &      &         &     section.c:936 & T.O. &  T.O. &T.O. &T.O. &T.O. &T.O. &T.O. & 862m  \\
		83 &      &         &     bfd.c:1108  & T.O. & 983m & 361m & 1288m &  T.O. & 812m & 621m &  423m  \\
		84 &      &          &     bfdio.c:262  & T.O. & T.O. & 1123m & T.O.  & T.O.  & 1011m  & 833m & 712m  \\
		85 &      &         &     stabs.c:372 & T.O. &  T.O. &T.O.  &  T.O. &T.O.&  T.O. &T.O. &T.O.  \\ \hline
		86 & \multirow{5}{*}{ffmpeg} &  \multirow{5}{*}{4.0.1}   &     rawdec.c:268   &  T.O.  & T.O. &N/A  & N/A & N/A & T.O. & T.O. & 286m  \\
		87 &      &         &     decode.c:557  & T.O. &  T.O.  &N/A  & N/A &N/A & T.O. & T.O.  & 369m \\
		88 &      &         &     dump.c:632  & T.O. &  T.O.  &N/A  & N/A &N/A & 836m & 411m  & 139m \\
		89 &      &         &     utils.c  & T.O. &  T.O.  &N/A  & N/A &N/A & T.O. & T.O.  & 863m \\
		90 &      &         &     eatgv.c:274 & T.O. & T.O. &N/A  & N/A &N/A & T.O. & T.O. &  1021m  \\ \hline
		91 & \multirow{5}{*}{mujs} &   \multirow{5}{*}{1.0.2}   &     jsrun.c:572  & T.O. & T.O.  &N/A & T.O.&N/A  & T.O.& T.O.& T.O.\\
		92 &      &         &    jsgc.c:47 & T.O. & T.O.  &N/A & T.O.&N/A  & T.O.& T.O.& T.O. \\
		93 &      &         &     jsdump.c:292 & T.O. & T.O.  &N/A & T.O.&N/A  & T.O.& T.O.& 652m \\
		94 &      &         &     jsvalue.c:362 & T.O. & T.O.  &N/A & T.O.&N/A  & 1013m & 736m & 539m \\
		95 &      &         &     jsvalue.c:396  & T.O. &   1165m  &N/A & 968m &N/A & 761m & 561m & 523m \\ \hline
		96 & \multirow{5}{*}{swftools} &   \multirow{5}{*}{0.9.2}    &     initcode.c:242  & 324m &301m &N/A & 223m  & N/A &241m & 131m &   89.3m  \\
		97 &      &        &     png.c:410   & 871m & 769m  &N/A  & 681m & N/A &617m &433m &  364m \\
		98 &      &        &     poly.c:137  & T.O. & T.O.  & N/A & T.O. & N/A  & T.O. & T.O.& T.O. \\
		99 &      &        &     jpeg2swf.c:257 & 677m  &632m & N/A & 541m & N/A &541m  &484m &355m \\
		100 &      &         &    swfc.c:1794  & T.O. & T.O. & N/A &  T.O.  & N/A& T.O.& T.O.& T.O. \\
		
		\midrule
		\multicolumn{4}{c}{\fontsize{6}{5}\selectfont \textbf{speedup}}  & {\textbf{37.75$\times$}} & {\textbf{29.11$\times$}} & {\textbf{23.34$\times$}} & {\textbf{95.61$\times$}} & {\textbf{143.22$\times$}} & {\textbf{23.01$\times$}}& {\textbf{14.80$\times$}} &{\textbf{-}} \\ 
		\multicolumn{4}{c}{\fontsize{6}{5}\selectfont \textbf{mean $\hat{A}_{12}$}} & {\textbf{0.88}} & {\textbf{0.85}} & {\textbf{0.92}} & {\textbf{0.86}} & {\textbf{0.91}} & {\textbf{0.82}} & {\textbf{0.79}} &{\textbf{-}} \\
		\multicolumn{4}{c}{\fontsize{6}{5}\selectfont \textbf{mean p-values}} & {\textbf{$0.002$}} & {\textbf{$0.008$}} & {\textbf{$0.008$}} & {\textbf{$0.003$}} & {\textbf{$0.001$}} & {\textbf{$0.009$}} &{\textbf{$0.012$}} &{\textbf{-}} \\  
		\bottomrule[1.5pt]
		
	\end{tabular}
	\begin{tablenotes}
		\tiny
		\item{*}  
		T.O. means that the fuzzers cannot reach target sites within 24 hours and N/A means that the fuzzer cannot successfully test the programs
	\end{tablenotes}
\end{table*}
\subsection{The Impact of the Optimizations on the Overall Performance}
\label{section:effective}
\noindent To answer RQ3,  
we conducted incremental experiments to evaluate the effects of the three optimizations on HyperGo's overall performance.
\begin{figure}[b]

	\setlength{\abovecaptionskip}{0.1cm}
	\centering
	\noindent \includegraphics[width=\columnwidth]{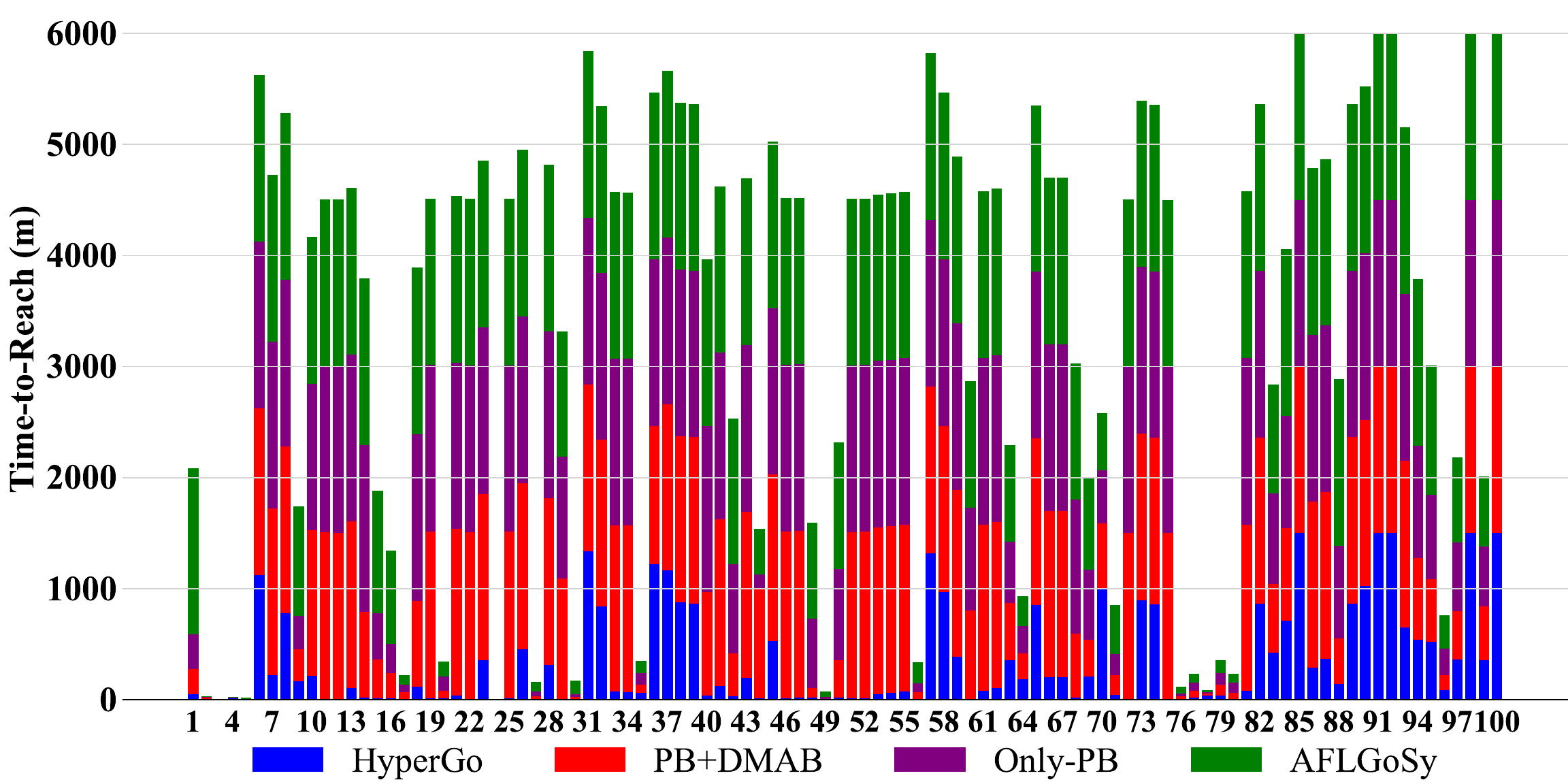}
	\caption{\label{fig:6}Incremental experiment results of AFLGoSy, Only-PB, PB+DMAB, and HyperGo using TTR.}
	\vspace{-0.3cm}
\end{figure}  
We use AFLGoSy as the base tool.
Since the DMAB model and OSEC scheme are based on the probability-based distance, disabling the probability-based distance calculation module will disable the other modules. Thus, we first add the probability-based distance module to AFLGoSy to implement a new tool (i.e., Only-PB). Then, we add the DMAB model to Only-PB, forming a new tool (i.e., PB+DMAB). Finally, we add the OSEC scheme to PB+DMAB to form HyperGo. In the incremental experiment, the configurations and the target sites remain unchanged as Section \ref{reach}.

According to the TTR results, Only-PB (41), PB+DMAB (46), and HyperGo (95) can all reach more target sites than AFLGoSy (38). Moreover, HyperGo outperforms AFLGoSy, Only-PB, and PB+DMAB by 33.79$\times$, 23.01$\times$, and 14.78$\times$ respectively in the average TTR of reaching the target sites. Detailed results are listed in Table \ref{table:unibench}. These results demonstrate that \textbf{each optimization has a significant impact on reducing TTR, and using one or two optimization strategies (Only-PB and PB+DMAB) are far less effective than using all three optimization strategies simultaneously (HyperGo)}. 
To visualize the experimental results, the results of TTR are shown in Figure \ref{fig:6}, in which the x-axis represents the target site ID (1-100), and the y-axis represents the total TTR of all fuzzers in minutes. 

\subsection{Intermediate data analysis} 

To demonstrate that HyperGo is more accurate than static-based DGF techniques and to more intuitively illustrate the effects of different optimizations, we analyzed the intermediate experimental data and used three metrics for analysis: 

(1) The number of reachable seeds generated by the fuzzers, i.e., \textbf{Rseeds}. Through Rseeds, we can observe whether a fuzzer can cover more paths leading to target sites, thereby reflecting a fuzzer's accuracy in analyzing path reachability and the capability of satisfying path constraints.

(2) The proportion of reachable seeds (i.e., \textbf{PRseed}) among all seeds. If the number and proportion of reachable seeds are higher, it indicates that the fuzzer can avoid spending time on infeasible and unreachable paths. 

(3) The number of reachable seeds generated by the symbolic executor, i.e., \textbf{SRseeds}. The more SRseeds indicate that the symbolic executor can provide more assistance to the fuzzer to cover new paths.

We evaluated the fuzzers on the programs from UniBench and counted the number of these three seed types, which are presented in Table \ref{table:seeds}.   	    	

\begin{table}[b]
	\vspace{-0.3cm}
	\footnotesize
	\centering
	\setlength{\abovecaptionskip}{0.1cm} 
	\setlength{\belowcaptionskip}{0cm}
	\setlength{\tabcolsep}{1.5pt}
	\caption{Intermediate data analysis using different seeds}
	\label{table:seeds}
	\begin{tabular}{ccccccccc}
		\toprule[1.5pt]
		&AFLGo & {BEAC} & {Wind}  & {Parm} &{AFSy} &{On-PB} & {PB+DM} & {HyperGo} \\ \midrule
		RSeeds & 2311 & 312 & 2532 & 1463 & 2479 & 5112 & 7313 & 12432\\
		PRseeds & 52.4\% & 79.8\% & 64\%  & 43.8\% & 
		46.3\% & 63.1\% & 69.3\% & 78.2\% \\
		SRseeds   & - & - & - & - & 18 & 22 & 29 & 267 \\
		
		\bottomrule[1.5pt]
	\end{tabular}
\end{table}
From Table \ref{table:seeds}, we can draw two conclusions. \textbf{Firstly, HyperGo can more accurately and efficiently explore more reachable and feasible paths to the target sites compared with other directed greybox fuzzers.}
By comparing the Rseeds of all fuzzers, we can see that HyperGo can generate more reachable seeds within the same time budget. Although BEACON has higher PRseeds, it has the lowest Rseeds among all fuzzers due to wrongly pruning some reachable paths. The inaccuracy of static analysis prevents BEACON from exploring more reachable paths to the target sites. Apart from BEACON, HyperGo has the highest PRseeds among all fuzzers. 
\textbf{Secondly, the performance of HyperGo, which uses all three optimizations, is significantly better than that of the fuzzers using only one (Only-PB) or two strategies (PB+DMAB)}. For the average number of RSeeds, both Only-PB (5112) and PB+DMAB (7313) are much greater than that of AFLGoSy (2479). This indicates that the probability-based distance and DMAB model can effectively explore more reachable paths to the target sites. 
As for the SRseeds, those of HyperGo are much greater than those of other fuzzers. This indicates that the OSEC scheme significantly improved the efficiency of symbolic execution, which helped and tested code areas that are difficult for the fuzzer to reach. 
Furthermore, according to the number of all three seed types, we can see that the performance of HyperGo, which uses all three optimizations, is significantly better than that of the fuzzers using only one or two strategies. This implies that the overall design of HyperGo, including the new fitness metric, optimized power schedule, and the OSEC scheme, works in a complementary way and achieves significant improvement.   
\subsection{Branch probability distribution analysis of the unexplored branches} 
\label{section:dist}

\noindent In the incremental experiments and intermediate data analysis, we observe that HyperGo can reach target sites faster than AFLGoSy and explore more paths leading to target sites. To verify whether the seeds preferentially selected by HyperGo using probability-based optimizations are better than those non-probability tools (RQ4), we analyze the high-priority seeds generated by HyperGo and AFLGoSy when testing UniBench. 
First, we divide the fuzzing process into 24 intervals, and each lasts for one hour. At each time point, such as 1-hour, we collect the top 100 seeds with the highest priority from the programs under test. 
Hence, at each time point, we collect 1600 seeds with the highest priority from the programs of UniBench. Then, we analyze the probability distribution of the seeds' unexplored branches. 

\begin{figure}[b]
\setlength{\abovecaptionskip}{0.15cm}

\begin{flushleft}
\subfigtopskip=2pt
\subfigbottomskip=-1pt
\subfigure[HyperGo \label{fig:hypergo}]{
	\includegraphics[width=4cm]{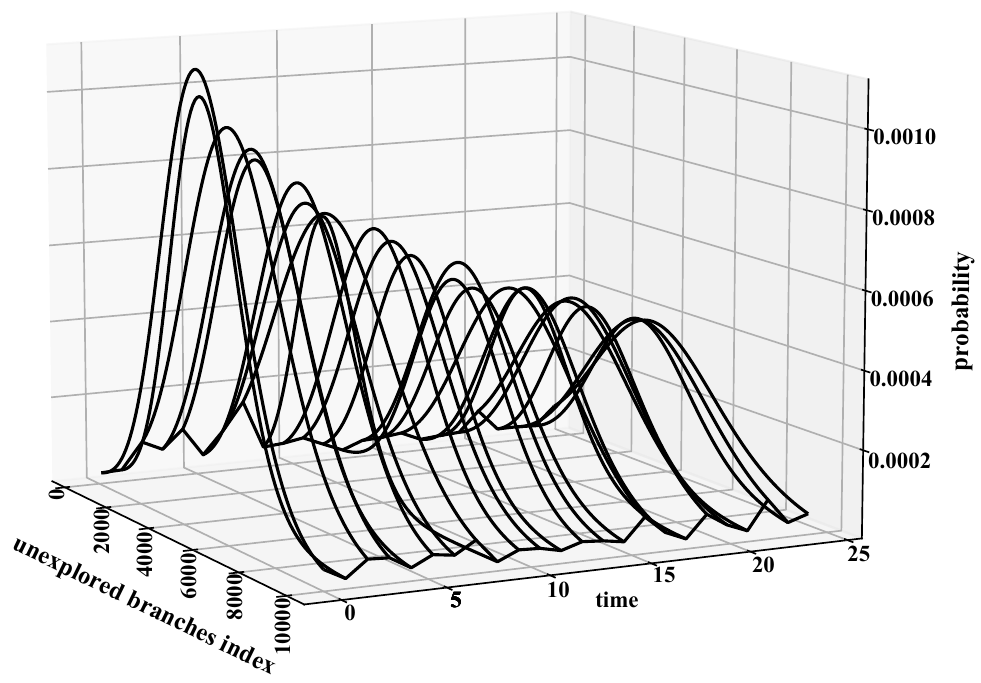}
}%
\hspace{-2mm}
\subfigure[AFLGosy \label{fig:aflgo}]{
	\includegraphics[width=4cm]{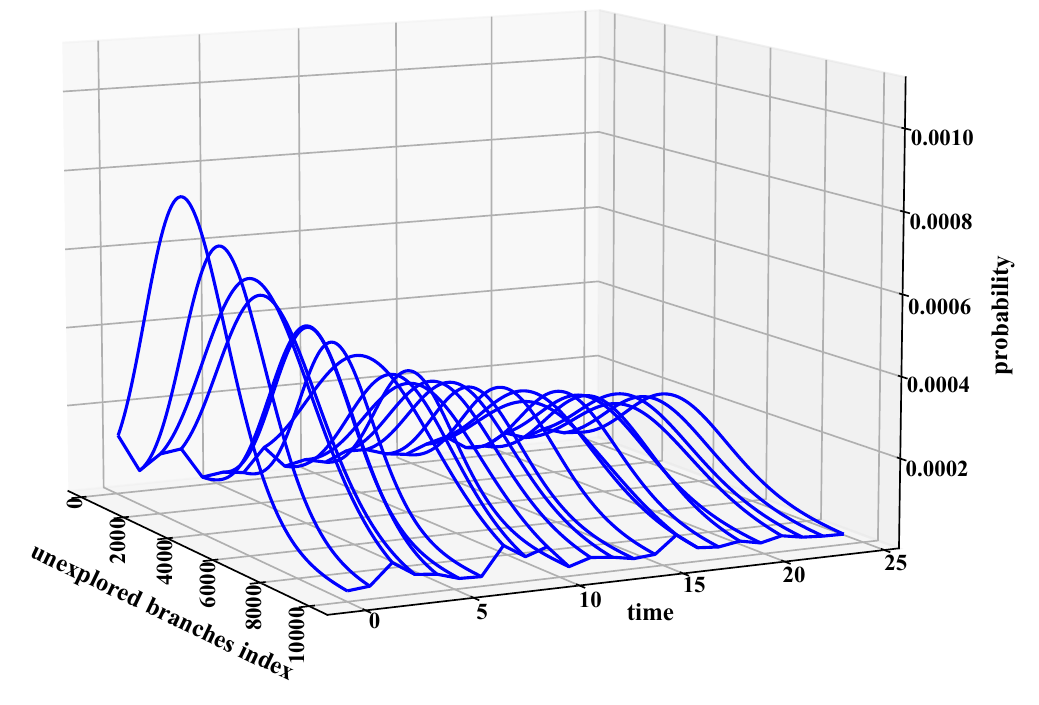}
}

\caption{\label{fig:proba}  Branch probability distribution of HyperGo and AFLGoSy.} 
\label{distribution}
\end{flushleft}
\end{figure}

The results are shown in Fig. \ref{fig:proba}, which includes two subfigures, depicting the unexplored branch probability distribution for HyperGo and AFLGoSy, respectively. For each subfigure, the x-axis represents the index of unexplored branches, the y-axis represents different time points, and the z-axis represents the branch probability. Each point (x, y, z) on the coordinate axis represents the branch probability of the $x^{th}$ unexplored branch at $y^{th}$ time point. Both Fig. \ref{fig:hypergo} and Fig. \ref{fig:aflgo} contain 24 branch probability distribution curves, corresponding to the 24 time points. To better illustrate the results, we sorted all branch probabilities and placed unexplored branches with higher probabilities closer to the middle. 
Then, to better visualize the overall branch probability, we performed curve fitting on all branch probabilities, resulting in a curve resembling a Gaussian distribution. 
We can obtain the maximum branch probability of different time points by identifying the highest point in the middle of the curve, and we can evaluate the overall branches' probabilities according to the overall height of the curve.

From Fig. \ref{fig:proba}, we can observe that the height of the AFLGoSy and HyperGo curves decreases gradually with the increasing of time, indicating the branch probability of all unexplored branches decreases during the fuzzing process. This is because the fuzzer is getting difficult to satisfy the branch conditions of unexplored branches as the fuzzing iterations increase.
Furthermore, it is noteworthy that HyperGo's unexplored branches generally exhibit higher branch probabilities than AFLGoSy's. This indicates that 
the overall seed quality of HyperGo is better than that of AFLGoSy, making HyperGo can explore more paths toward target sites.
This can be attributed to HyperGo's probability-based optimizations, which prioritize seeds that are more likely to cover unexplored branches as compared to AFLGoSy. Hence, we can conclude that \textbf{the probability-based fitness metric employed by HyperGo plays a crucial role in discovering better seeds, which are more likely to cover unexplored branches so as to explore more paths toward target sites}. This is essential for achieving faster attainment of the target sites in directed fuzzing.

\subsection{Discovering new vulnerabilities}
\label{section:new}

To answer RQ4, we used HyperGo to test real-world programs. We first used sanitizers (i.e., UBSAN \cite{53} and ASAN\cite{54}) to locate and label potential vulnerabilities as the target sites. Then, we run HyperGo for 24 hours to detect new vulnerabilities. Finally, HyperGo discovered 10 undisclosed vulnerabilities from 5 real-world programs. The information about these vulnerabilities is presented in Table \ref{table:10}. From the table, we can see that the new vulnerabilities involve heap-buffer-overflow, out-of-bounds read/write, and Null pointer deference.

We also used the baseline fuzzers to detect them. As a result, among the 10 discovered vulnerabilities, 5 could also be detected by AFLGo, 5 by AFLGoSy, 5 by BEACON, 4 by WindRanger, and 2 by ParmeSan. As for the reason that the vulnerabilities could not be detected, one is that the fuzzer could not run the program or obtain the distance information for analysis, while the other reason is the vulnerability could not be triggered within the time budget. 
From the above result, we can conclude that \textbf{HyperGo can detect new vulnerabilities from real-world programs, and it outperforms the baseline fuzzers}.
\begin{table*}[t]

	\centering
	\setlength{\abovecaptionskip}{0.1cm} 
	\setlength{\belowcaptionskip}{0cm}
	\setlength{\tabcolsep}{6pt}
	\caption{New vulnerabilities detected by HyperGo}
	\label{table:10}
	\begin{threeparttable}
		\begin{tabular}{cccccc}
			\toprule[1.5pt]
			\textbf{No} & \textbf{Prog} & \textbf{Bug location} & \textbf{Bug Type}
			& CNNVD-ID&\textbf{ GSBWPH} \\                
			\midrule
			1  &  cflow1.6 &  symbol.c:302   & heap-buffer-overflow  
			& 2023-88222684
			& $\times$ $\times$\checkmark\checkmark $\times$$\checkmark$  \\ 
			2  & 
			gdk-pixbuf-2.31     &  gdk-pixdata.c:439    &heap-buffer-overflow& 2023-38595027 &
			\checkmark \checkmark \checkmark \checkmark \checkmark \checkmark
			\\
			3  &  gdk-pixbuf-2.31    &  io-qtif.c:437     & out-of-bound read &2023-61429059
			&\checkmark\checkmark\checkmark\checkmark$\times$\checkmark
			\\
			4  &   gdk-pixbuf-2.31   &    io-pcx.c:271      &  heap-buffer-overflow  & 2023-36676426
			&$\times$$\times$$\times$$\times$$\times$\checkmark\\
			5  &  gdk-pixbuf-2.31     &   io-pcx.c:528       &  heap-buffer-overflow  & 2023-93057825 &\checkmark  \checkmark\checkmark \checkmark$\times$\checkmark\\
			6  & gdk-pixbuf-2.31  &  gdk-pixdata.c:142     &   heap-buffer-overflow &2023-18623971   & $\times$$\times$\checkmark$\times$$\times$\checkmark\\ 
			
			7 &   jhead-3.00    & jpgqguess.c:195  & heap-buffer-overflow    & 2023-28389092
			& $\times$$\times$$\times$$\times$\checkmark\checkmark\\
			
			8 &     	flvmeta-1.2.1  &      dump\_xml.c:271 & out-of-bound read & 2023-88566232 &$\times$$\times$$\times$$\times$$\times$\checkmark\\
		
			9 &fig2dev     &bound.c:525 &Null pointer dereference& 2023-43290258&\checkmark\checkmark $\times$$\times$$\times$\checkmark \\
			10 &fig2dev     &arrow.c:89 &out-of-bound read & 2023-87146636&\checkmark\checkmark $\times$$\times$$\times$\checkmark \\

			\bottomrule[1.5pt]
		\end{tabular}
	\end{threeparttable}
	\begin{tablenotes}
		\footnotesize
		\item{1*} In the last column, letters G,S,B,W,P, and H represent AFLGo, AFLGoSy, BEACON, WindRanger, ParmeSan, and HyperGo, respectively. 
		\item{2*} `$\times$' denotes that the fuzzer was unable to discover the vulnerability, while `\checkmark' signifies that the fuzzer was able to discover the vulnerability.
	\end{tablenotes}
	\vspace{-0.2cm}
\end{table*}
We use the example in Listing \ref{lst3} as a case study to explain why HyperGo could discover more vulnerabilities than the baseline fuzzers. Listing \ref{lst3} shows a heap-overflow vulnerability in function \verb:read_pixel_4(): of gdk-pixbuf 2.31.1. At Line 16, due to the lack of range restriction on variable \texttt{offset}, if the value of \texttt{offset} exceeds the memory allocated for the array \texttt{data}, a heap overflow would occur. To trigger this vulnerability, the fuzzer needs to generate inputs that satisfy both the path constraints at Line 1
\begin{lstlisting}[caption={Example of a heap-overflow in gdk-pixbuf 2.31.1.},captionpos=b,label=lst3,style=mystyle]
static gboolean gdk_pixbuf__pcx_load_increment{
 if(context->current_task==PCX_TASK_LOAD_DATA) {
  switch(context->bpp) {
   ...
   case 4:
    retval=pcx_increment_load_data_4(context);
    static gboolean pcx_increment_load_data_4(){
     ...
     @p=read_pixel_4(planes[0], i)&0xf;@
    }
   }
 }
}
static guchar read_pixel_4(){
 if(!(offset % 2))
  @etval = data[offset] >> 4;@
}
\end{lstlisting}
  and Line 3 and satisfy the root cause of the vulnerability. Within the time budget of 24 hours, AFLGo, AFLGoSy, WindRanger, and BEACON failed to generate specific inputs that satisfy all three conditions simultaneously to trigger this vulnerability. Based on three optimization strategies, HyperGo was able to reach Line 10 more efficiently and generated an input that triggers the vulnerability within 160 minutes.

\section{Discussion}
HyperGo adopts three optimizations to enhance the directedness, including the probability-based distance, the DMAB model, and the OSEC scheme. 
Specifically, the probability-based distance prioritizes the optimal seeds which have shorter seed distances and higher path probabilities. The DMAB model optimizes the power schedule, which implicitly balances the exploitation of seeds with short distances and the exploration of more reachable seeds.  
The OSEC scheme combines DGF and SE in a complementary manner. After pruning the unreachable and unsolvable branches, HyperGo prioritizes the symbolic execution of the seeds with higher scores to accelerate the speed of reaching targets. 
Experiments have proved the effectiveness of HyperGo in reaching the target sites (Section \ref{reach}), exposing the known vulnerabilities (Section \ref{expose}), and discovering new vulnerabilities (Section \ref{section:new}). Moreover, we also proved the effectiveness of the three optimizations (Section \ref{section:effective}), and we can visually see their effectiveness via branch probability distribution of unexplored branches (Section \ref{section:dist}).

Different from the SOTA experience-based and intuition-based DGF techniques, HyperGo adopts the probability-based fitness metrics and improvement methods that allowed it to maintain high accuracy across testing different programs in different fuzzing phases. The probability is calculated according to the simple branch hits rather than relying on program analysis or expert knowledge. Therefore, the probability-based fitness metric, the OSEC scheme, and the DMAB model can adaptively select the optimal seeds or optimal paths in current fuzzing phases according to the testing information. Compared to other DGF techniques, HyperGo's adaptability allows it to have higher accuracy when testing most programs.

\textbf{Threat to validation}. Aiming to design an adaptive approach, we adopt several heuristic parameters in HyperGo, which are set empirically (e.g., the setting of the adjustment factor). The values of these parameters might to some extent affect the performance of HyperGo. However, after extensive experiments, we believe the setting of these parameters is stable and suitable for most of the testing scenarios.

\section{Related Work}
In this section, we focus on discussing the most related works: directed greybox fuzzing and directed hybrid fuzzing.

\textbf{Directed Grey-box Fuzzing.}
AFLGo is the first directed greybox fuzzer. It calculates the distances between the seeds and pre-defined targets to prioritize the seeds closer to the targets, which casts reachability as an optimation problem to minimize the distance between the seeds and their targets. Based on AFLGo's idea, Hawkeye \cite{21} proposes the concept of trace similarity and adjusts its seed prioritization, power scheduling, and mutation strategies to enhance directedness. However, Hawkeye suffers the same issues as those of AFLGo when encountering complex path constraints. Even if they assign more energy to the closer seeds, it is difficult for them to satisfy the complex path constraints to cover the path toward target sites.                        
Some directed greybox fuzzers, such as LOLLY \cite{60}, Berry \cite{berry}, UAFL \cite{58}, and CAFL \cite{22}, propose new fitness metrics, such as sequence similarity, to enhance directedness and detect hard to manifest vulnerabilities. These methods are derived from the analysis of program characteristics or the root causes of different vulnerabilities. Thus, for some specific programs or fuzzing processes, these new fitness metrics may be inaccurate and take a negative effect, which has been discussed in Section \ref{motivation}.
Other directed greybox fuzzers use data flow information and data conditions information to enhance directedness. WindRanger \cite{23} uses the deviation basic blocks (DBBs) and the data flow information for seed distance calculation, seed mutation, seed prioritization, and power schedule. BEACON \cite{26} leverages a provable path-pruning method to reduce the exploration of infeasible paths. 
However, due to the limitations of static analysis, BEACON's analysis of infeasible paths (e.g., BEACON cannot recognize indirect calls) may be inaccurate. This can result in the incorrect pruning of some feasible paths, and consequently slowing down the process of reaching target sites.
Besides, FuzzGuard \cite{27} uses the deep neural network to extract the features of reachable seeds and filter out the unreachable seeds to improve efficiency. To search the inputs that can reach the target sites, $MC^2$ designs an asymptotically optimal randomized directed greybox fuzzer that has logarithmic expected execution complexity in the number of possible inputs.
However, DGF still suffers from being difficult to penetrate through the hard-to-satisfy path constraints. HyperGo selects the better paths that have fewer hard-to-satisfy path constraints and utilizes symbolic execution to assist DGF to pass through such path constraints.

\textbf{Directed Hybrid Fuzzing.} Directed Hybrid Fuzzing uses the heuristic strategies in hybrid fuzzing to gain directedness. 
Directed hybrid fuzzers achieve directedness by prioritizing the symbolic execution of reachable seeds or closer seeds. Hydiff \cite{36}, SAVIOR \cite{28} and Badger\cite{30} prioritize the seeds that may cause the specific program bug locations as the target sites, and then prioritizes symbolic execution of the seeds which are reachable from more target sites. DrillerGO \cite{16}, 1dvul\cite{1dVul}, and Berry \cite{berry} combine the precision of DSE and the scalability of DGF to mitigate their individual weaknesses. However, modern directed hybrid fuzzers suffer from the limitation of symbolic execution. Since the symbolic executor may fail to solve many unexplored branches or succeed in solving the unreachable branches, such useless constraint solving will have a negative impact on the directedness of directed hybrid fuzzing. Thus, HyperGo uses the OSEC scheme to prune the unreachable and unsolvable branches and prioritize the symbolic execution of the optimal seeds to better combine DGF and SE.

\vspace{-0.15cm}
\section{Conclusion}

In this paper, we propose HyperGo, a probability-based directed hybrid fuzzer.
HyperGo adopts the probability-based distance as the fitness metric and an optimized power schedule (namely DMAB model), which can steer DGF to faster reach the target sites through the paths that are easier to re-exercise and closer to the target sites. Using the OSEC scheme, HyperGo combines DGF and SE in a complementary manner to focus on solving constraints toward reachable targets.
HyperGo is evaluated on 100 target vulnerabilities of 21 real-world programs from 2 datasets, the experiment results show that HyperGo outperforms the state-of-the-art directed fuzzers (AFLGo, BEACON, WindRanger, and ParmeSan) in reaching target sites and exposing known vulnerabilities. Moreover, HyperGo also discovered 10 undisclosed vulnerabilities and demonstrated its effectiveness in vulnerability discovery.

\section*{Acknowledgments}
This work is partially supported by the National Key Research and Development Program of China under Grant No. 2021YFB0300101, the National Natural Science Foundation China (62272472, 61902405, U22B2005, 61972412, 62306328), the HUNAN Province Natural Science Foundation (2021JJ40692), and the National High-level Personnel for Defense Technology Program (2017-JCJQ-ZQ-013).

\bibliographystyle{ACM-Reference-Format}
\bibliography{HyperGo}

\end{document}